\documentclass[nc, reprint,showpacs, superscriptaddress]{revtex4-1}

\usepackage{color,xcolor}

\usepackage{footnote}
\usepackage[colorinlistoftodos,bordercolor=red!50,backgroundcolor=red!10,linecolor=red!50,textsize=tiny]{todonotes}
\usepackage{changes}
\usepackage[utf8]{inputenc}
\usepackage[T1]{fontenc}
\usepackage[english]{babel}  % English language
\usepackage{times}
\usepackage{bm}
\usepackage{ulem}

\usepackage[colorlinks=true,linkcolor=blue,citecolor=blue,urlcolor=blue,pdfauthor={ },pdftitle={ },pdfsubject={ },pdfkeywords={ }]{hyperref}
\usepackage{amssymb,amsmath,amsfonts,amsthm}
\usepackage{mathtools}
\usepackage{verbatim}
\usepackage{enumerate}
\usepackage{graphicx}
\usepackage{hyperref}
\usepackage{bbm}
\usepackage{booktabs}

\usepackage[caption=false]{subfig}

\usepackage{setspace}
\usepackage{array}

\usepackage{color}
\definecolor{dred}{rgb}{.8,0.2,.2}
\definecolor{ddred}{rgb}{.8,0.5,.5}
\definecolor{dblue}{rgb}{.2,0.2,.8}
\definecolor{dgreen}{rgb}{.2,0.5,.2}

\newcommand{\bra}[1]{\mbox{$\langle #1|$}}
\newcommand{\ket}[1]{\ensuremath{|#1\rangle}}

\newcommand{\be}{\begin{equation}}
\newcommand{\ee}{\end{equation}}
\newcommand{\bea}{\begin{eqnarray}}
\newcommand{\eea}{\end{eqnarray}}

\begin{document}

\title{Optimal Quantum Overlapping Tomography}

\author{Chao Wei}
\affiliation{Shenzhen Institute for Quantum Science and Engineering, Southern University of Science and Technology, Shenzhen 518055, China}
\affiliation{International Quantum Academy, Shenzhen 518048, China}
\affiliation{Guangdong Provincial Key Laboratory of Quantum Science and Engineering,
Southern University of Science and Technology, Shenzhen 518055, China}
\affiliation{Shenzhen Key Laboratory of Quantum Science and Engineering, Southern University of Science and Technology, Shenzhen,518055, China}

\author{Tao Xin}
\email{xint@sustech.edu.cn}
\affiliation{Shenzhen Institute for Quantum Science and Engineering, Southern University of Science and Technology, Shenzhen 518055, China}
\affiliation{International Quantum Academy, Shenzhen 518048, China}
\affiliation{Guangdong Provincial Key Laboratory of Quantum Science and Engineering,
Southern University of Science and Technology, Shenzhen 518055, China}
\affiliation{Shenzhen Key Laboratory of Quantum Science and Engineering, Southern University of Science and Technology, Shenzhen,518055, China}

%%%%%%%%%%%%%%%%%%%%%%%%%%renyi+Pn+coherence(no negativity
\begin{abstract}
Partial tomography, which focuses on reconstructing reduced density matrices (RDMs), has emerged as a promising approach for characterizing complex quantum systems, particularly when full state tomography is impractical. Recently, overlapping tomography has been proposed as an efficient method for determining all $k$-qubit RDMs using logarithmic polynomial measurements, though it has not yet reached the ultimate limit. Here, we introduce a unified framework for optimal quantum overlapping tomography by mapping the problem to the clique cover model. This framework provides the most efficient and experimentally feasible measurement schemes to date, significantly reducing the measurement costs. Our approach is also applicable to determining RDMs with different topological structures. Moreover, we experimentally validate the feasibility of our schemes on practical nuclear spin processor using average measurements and further apply our method to noisy data from a superconducting quantum processor employing projection measurements. The results highlight the strong power of overlapping tomography, paving the way for advanced quantum system characterization and state property learning in the future. 

  \end{abstract}

\maketitle
%%%%%%%%%%%%%%%%%%%%%%%%%%%%%%%%%%%%%%%%%%%%%%%%%%%%%
\textit{Introduction.—} 
Quantum state tomography (QST), which involves reconstructing the microscopic wave function, is a fundamental task in modern quantum information science. In practice, we often perform full QST to extract results from quantum information processing, evaluate the quality of experimentally prepared states, or estimate specific properties of states \cite{james2001measurement,merkel2013self,eisert2020quantum}. However, the exponential increase in the number of required measurements makes it impractical for large systems \cite{cramer2010efficient,flammia2012quantum,lanyon2017efficient,huang2020predicting}. For example, reconstructing a ten-qubit state can take around five days \cite{song201710}. 

Recently, numerous studies have shown that these tasks can be performed more efficiently by using partial tomography on $k$-qubit reduced density matrices ($k$-RDMs). For instance, full states can be determined and reconstructed from their RDMs \cite{diosi2004three,chen2013uniqueness,xin2017quantum,guo2024quantum}. Additionally, properties such as state fidelity \cite{flammia2011direct,cerezo2020variational,wu2023quantum,qin2024experimental}, entanglement entropy \cite{islam2015measuring,elben2020mixed,huang2022measuring, wu2023learning}, physical Hamiltonian parameters \cite{wang2015hamiltonian,xin2019local,che2021learning}, and the energy in variational quantum eigensolvers \cite{peruzzo2014variational,verteletskyi2020measurement,tilly2022variational,claudino2020benchmarking,gupta2022variational} can be estimated from measurements on local observables that encompass all $k$-RDMs, often with the help of neural networks and tensor networks \cite{miranowicz2014optimal,melko2019restricted,koutny2022neural,carrasquilla2019reconstructing,akhtar2023scalable}. A direct and commonly used approach involves measuring each $k$-RDM independently. However, this method is inefficient, as measuring one $k$-RDM often provides information about other overlapping $k$-RDMs.

Building on this principle, quantum overlapping tomography (QOT) techniques are proposed for reconstructing all $k$-RDMs more efficiently \cite{cotler2020quantum, bonet2020nearly}. They suggest that single-qubit Pauli measurements are performed in parallel, where each qubit is measured in the Pauli basis $\{X, Y, Z\}$, such that all $k$-RDMs of a $N$-qubit state can be determined with a logarithmic number of parallel measurements. For $k=2$, they claim that $3+6\lceil \text{log}_2N \rceil$ parallel measurement settings are sufficient to determine all 2-RDMs, with generalizations to $k>2$ also being possible. This value is reduced to $3+6\lceil \text{log}_3N \rceil$ but without generalizations for $k>2$ in \cite{garcia2020pairwise}. Since its proposal, QOT has garnered significant experimental interest. It has been used to estimate state fidelity in nuclear spin platforms \cite{qin2024experimental}, determine 2-RDMs and entanglement entropy in optical platforms \cite{yang2023experimental}, and reconstruct full states in superconducting platforms \cite{guo2024quantum, hu2024experimental}. However, the limit of this scheme remain an unexplored problem.

In this letter, we address this problem and demonstrate the application of our scheme in experiments. Specifically, we develop a more efficient QOT by mapping the problem to that of a clique cover problem, which yields parallel measurement schemes with the minimum number of measurement settings required to determine all $k$-RDMs. The value $k$ is often referred to as the locality length of the RDM. Additionally, our approach enables the design of optimal measurement schemes for reconstructing RDMs with varying locality lengths that the current QOT cannot handle. Moreover, the experimental demonstrations on four-qubit nuclear spin and six-, nine-qubit superconducting quantum processors confirm the practical feasibility and advantage of our scheme, reducing both the number of measurement settings and samples.

 \begin{figure}
 \centering
  \includegraphics[width=0.49\textwidth]{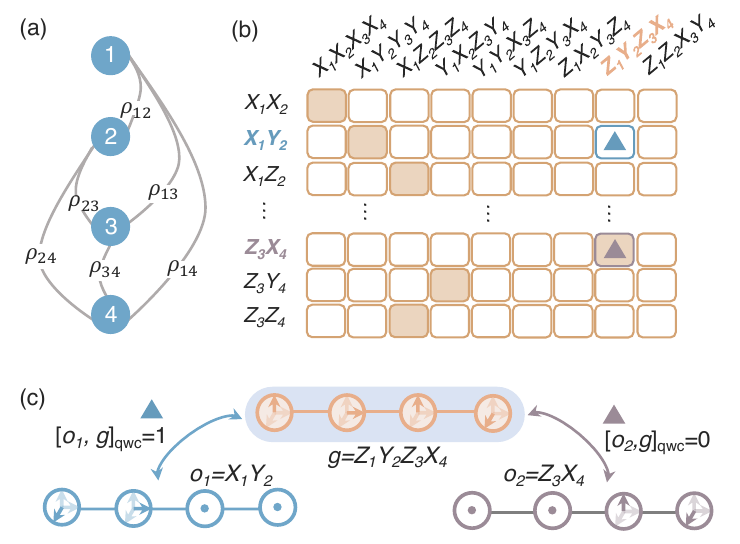}
   \caption{ Our overlapping tomography for determining all $k$-RDMs for the example $(4, 2)$. (a) There are ${N \choose k} =6$ two-qubit RDMs: $\rho_{12}, \rho_{13}, \rho_{14}, \rho_{23}, \rho_{24}$, and $\rho_{34}$ (represented by the gray lines). If each RDM is measured individually, a total of ${3^k}{N \choose k} =54$ local observables need to be measured. (b) The qubit-wise commutativity matrix shows that, according to our optimal QOT scheme, measuring just 9 parallel observables is sufficient to determine these 2-RDMs. (c) The qubit-wise commutativity between $\hat{o}_1=X_1Y_2$ and $\hat{o}_2=Z_3Y_4$ with the parallel observable $\hat{g}=Z_1Y_2Z_3X_4$ determines the element value of the matrix  $\mathcal{A}$. } 
  \label{fig1}
\end{figure}

\textit{Optimal QOT.—}  For a $N$-qubit state $\rho$, the $k$-RDMs $\rho_{k}=\text{tr}_{j\neq{i_1,...,i_k}}[\rho]$ are obtained by tracing out all qubits except those indexed by $i_1,...,i_k$. $\rho_{k}$ can be expressed in the complete Pauli basis $\{I, X, Y, Z\}^{\otimes k}$, 
\begin{equation}
\rho_{k}=\frac{1}{2^k}\sum_{j, ..., l=0}^3\text{tr}[\sigma_j^{(1)}\otimes...\sigma_l^{(k)}\rho_{k}] \sigma_j^{(1)}\otimes...\sigma_l^{(k)},
\end{equation}
with $\sigma_{0,1,2,3}=I, X, Y, Z$. $\rho_{k}$ is typically determined by performing partial tomography on the system, requiring the measurement of $4^k-1$ expectation values, where `$-1$' arising from the unit trace condition. Single-qubit measurements are typically performed on each qubit in practical $k$-qubit measurements, allowing the reuse of these measurements to reconstruct other expectation values. For example,  the expectation values of $X_1$ and $X_2$ can be inferred from the measurements of the observables $X_1X_2$ when reconstructing the RDM $\rho_{12}$. Thus, only $3^k$ Pauli observables need to be measured to reconstruct each $k$-qubit RDM. There are ${N \choose k} $ such $k$-RDMs in the ($N$,$k$) case, as illustrated in Fig. \ref{fig1}(a). Therefore, measuring ${3^k}{N \choose k} $ observables is required to reconstruct all $k$-RDMs. However, this straightforward approach is obviously inefficient, as it fails to account for the overlaps between different RDMs. 
 
Parallel measurements, where each qubit is measured in the basis $\{X, Y, Z\}$ in parallel, can significantly enhance the efficiency of reconstructing all $k$-RDMs. For example, measuring the parallel observable $X_1X_2X_3$ not only provides the expectation values of $X_1$, $X_2$, and $X_1X_2$ for reconstructing $\rho_{12}$, but also those of $X_1$, $X_3$, and $X_1X_3$ for reconstructing $\rho_{13}$, as well as the expectation values of $X_2$, $X_3$, and $X_2X_3$ for reconstructing $\rho_{23}$. Recently, some QOT techniques have been proposed for designing parallel measurements that can efficiently recover all $k$-RDMs, utilizing partitioning and the family of perfect hash functions \cite{cotler2020quantum, bonet2020nearly,garcia2020pairwise}. Here, we develop more flexible schemes which can provide the more efficient solutions to address this problem. Before that, we define a set $\mathcal{L}$ of local observables which comprise all $k$-RDMs, such as $\mathcal{L}=\{\sigma_{i_1}\otimes \sigma_{i_2}|\sigma \in [X, Y, Z]; i_1, i_2 \in [1, 2, ..., N]\}$ for 2-RDMs. Additionally, we define a set $\mathcal{G}$ of global observables that consists of parallel measurements on qubits. We use the qubit-wise commutativity $[\hat{o}, \hat{g}]_{\text{qwc}}$ to describe the relationship between $\hat{o}\in \mathcal{L}$ and $\hat{g}\in \mathcal{G}$ \cite{gokhale2019minimizing,yen2020measuring,verteletskyi2020measurement}. If each corresponding pair of one-qubit Pauli operators in $\hat{o}$ and $\hat{g}$ commutes, then $\hat{o}$ and $\hat{g}$ are qubit-wise commuting, and $[\hat{o}, \hat{g}]_{\text{qwc}} = 0$. Otherwise $[\hat{o}, \hat{g}]_{\text{qwc}}=1$. The key question now is how to determine the set $\mathcal{G}_{\text{opt}} \subseteq \mathcal{G}$ with the minimum elements that satisfies the following condition,
\begin{equation}
\exists~ \hat{g}\in \mathcal{G}_{\text{opt}},  [\hat{o}, \hat{g}]_{\text{qwc}}=0, \text{for} ~\forall ~\hat{o}\in \mathcal{L}.
\label{cod}
\end{equation}

Here, we map this problem to the clique cover model. First, we define a qubit-wise commuting matrix $\mathcal{A}$  that indicates whether the expectation values of $\hat{o}$ can be obtained from the measurements of $\hat{g}$.  If the $i$-th element $\hat{o}_i$ is qubit-wise commuting with the $j$-th element $\hat{g}_j$, meaning $[\hat{o}_i, \hat{g}_j]_{\text{qwc}}=0$ and measuring $\hat{g}_j$ provides the expectation values of $\hat{o}_i$, we define the matrix element $\mathcal{A}_{ij}=1$. Otherwise,  $\mathcal{A}_{ij}=0$. Figure \ref{fig1}(b-c) presents the qubit-wise commuting matrix for the example $(4, 2)$. Second, we define a binary vector $\mathbf{x}$.  An element $x_i=1$ indicates that the $i$-th parallel observable $\hat{g}_i$ is included in the set $\mathcal{G}_{\text{opt}}$. In this framework, the condition in Eq. (\ref{cod}) can be expressed as the simple constraint $\mathcal{A}\mathbf{x}\geqslant 1$. The number of parallel observables corresponds to the number of ones in the vector $\mathbf{x}$. In summary, the measurement scheme can be determined by solving the following problem, 
\begin{align}
\text{minimize:}  ~&f=||\mathbf{x}||, \\
~\text{subject to:}  ~&\sum_j\mathcal{A}_{ij}x_j\geqslant 1, x_i\in \{0,1\}. 
\label{qc}
\end{align}
with the vector norm $||\cdot||$. The problem can be addressed using binary linear programming optimization, which will yield the optimal parallel measurement design for recovering all $k$-RDMs. As shown in Fig. \ref{fig1}(b), our scheme demonstrates that 9 parallel observables are sufficient to reconstruct all 2-RDMs for $N=4$, while the current QOT method requires 15 parallel observables. Our approach offers several key improvements compared to previous QOT methods.

 \begin{figure}
 \centering
  \includegraphics[width=0.49\textwidth]{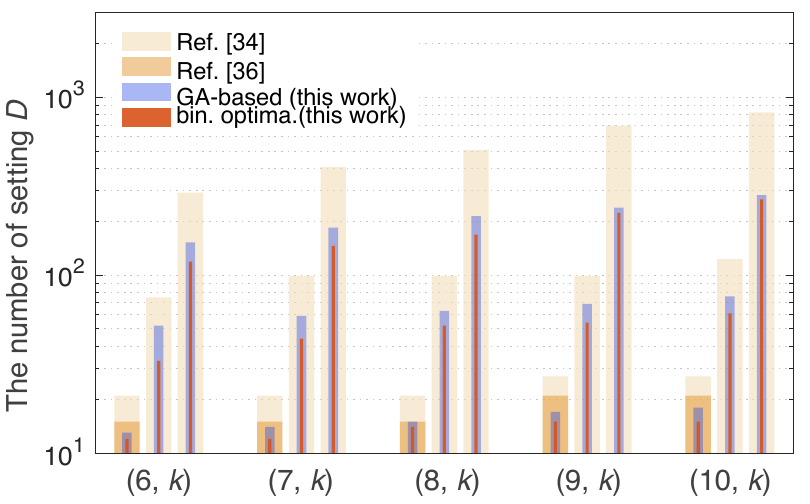}
  \caption{ The comparison between the results of our method and the previous QOT schemes. We compared the number of parallel observables required for $N=6\sim 10$ and $k=2\sim 4$ in different methods. Both QOT schemes in Ref. \cite{cotler2020quantum} and Ref. \cite{bonet2020nearly} yield identical results for $k=2$. However, Ref. \cite{cotler2020quantum} uses hash functions to generalize the scheme to higher $k$, achieving better results than Ref. \cite{bonet2020nearly}, so we present only the results from Ref. \cite{cotler2020quantum} for comparison. Although Ref. \cite{garcia2020pairwise} offers a superior measurement scheme compared to Refs. \cite{cotler2020quantum} and \cite{bonet2020nearly}, it is limited to $k=2$. In contrast, our work provides the most efficient measurement scheduling for arbitrary $k$ and surpasses all previous QOT methods.} 
  \label{fig2}
\end{figure}

First, our framework provides more efficient solutions for arbitrary $k$, including cases where $k > 2$. Figure \ref{fig2} shows the results for $N=6\sim10$ and $k=2\sim 4$, clearly demonstrating that our approach greatly enhances the QOT performance. It is highly beneficial for experimental tomography. Most current quantum processors can only perform measurements in the basis $\{I, Z\}^{\otimes N}$. It is necessary to design and implement readout operations before measurement in order to measure other Pauli bases. However, calibrating and compiling these readout operations in experiments often requires considerable physical time \cite{crawford2021efficient}. Therefore, reducing the number of parallel observables in our method will decrease the practical measurement resource costs.

Second, current QOT methods are effective only when the goal is to reconstruct all $k$-RDMs. However, in practical scenarios, the states often exhibit a topological structure, such as a one-dimensional chain, a two-dimensional lattice, or a center-spin model, where the focus is typically on nearest-neighbor $k$-RDMs rather than all $k$-RDMs. Our framework can also be adapted to these tasks by adjusting the target set of local observables. For example, our results show that $3^k$ parallel observables are the optimal design for nearest-neighbor $k$-RDMs in a one-dimensional chain, independent of the system size, which aligns well with previous empirical findings \cite{qin2024experimental, lanyon2017efficient,guo2023scalable}. For a two-dimensional lattice \cite{cao2023generation}, our results show that measuring 9 parallel observables is sufficient to determine nearest-neighbor 2-RDMs with any system size \cite{sm}. Furthermore, our method is also well-suited for reconstructing mixed RDMs with varying locality lengths, such as mixed RDMs composed of 2- and 3-RDMs \cite{sm}.

Moreover, the original QOT relies on constructing the family of perfect hash functions, which can be challenging due to the difficulty in minimizing the number of functions. Although our method employs the minimum clique cover, whose optimal solutions are also hard to obtain for larger systems, finding the exact optimal solutions is not necessary. There are many efficient algorithms for producing nearly optimal schemes. Here, we present the nearly optimal solution to this problem based on a scalable greedy algorithm \cite{sm}, as shown in Fig. \ref{fig2}, which still outperform the original QOT. 

 \begin{figure*}
 \centering
  \includegraphics[width=1\textwidth]{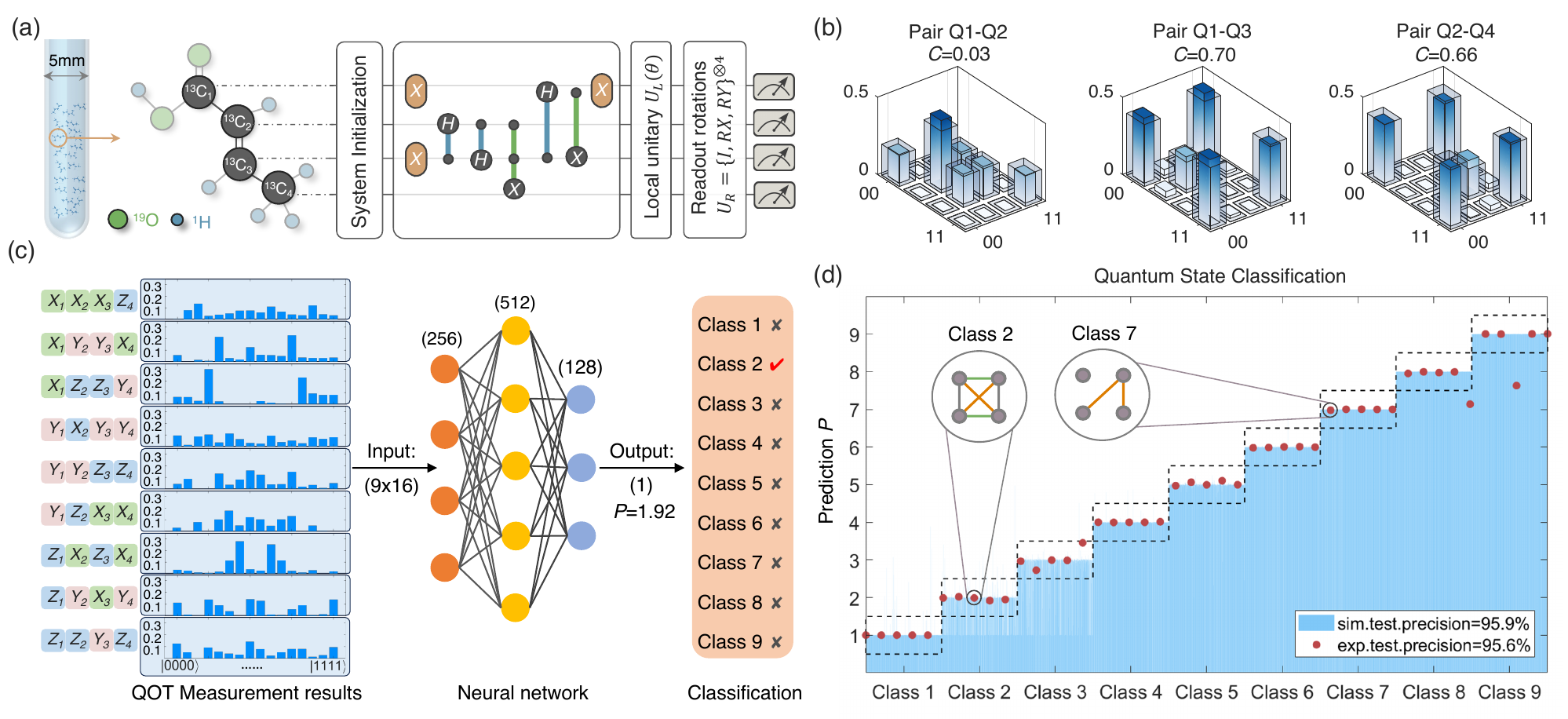}
   \caption{(a) The experimental setup containing the used samples with four nuclear spins and the state preparation and measurement circuits.  (b) The reconstructed 2-RDMs via our QOT method and the comparison with the ideal ones (transparent bars) for the state $\ket{\Psi_4}$. (c) The structure of our quantum state classifier. The measurement data on nine parallel observables is fed into the neural network for state classification. Here, we provide an example of the experimentally QOT measurement data alongside the prediction value $P=1.92$ (Class 2). (d) The classification results. We evaluate the classification performance using both numerically simulation data (blue bars) and experimentally measurement data (five states for each class). By setting a threshold window (dashed squares), we can classify the experimentally prepared states into different classes without the need for full QST. The states exhibit distinct concurrence patterns. For instance, the states in Class 2 demonstrate non-zero concurrence for every pair of qubits and display three different concurrence values represented by three colors of lines.}
  \label{fig3}
\end{figure*}

\textit{Experiments.—} To validate the feasibility and advantage of our method on practical quantum processors, we prepare a series of states, reconstruct their 2-RDMs, and further demonstrate the application of our QOT in quantum state classification on
nuclear magnetic resonance (NMR) with average measurements.  The sample consists of a bulk of molecules $^{13}$C-labeled trans-crotonic acid (C$_4$H$_6$O$_2$) dissolved in $d6$-acetone. Four carbons are used as a four-qubit processor after decoupling protons. The system Hamiltonian is written as, 
\begin{align}\label{Hamiltonian}
	\mathcal{H}_{sys}=-\sum\limits_{i=1}^4 {\pi \nu_i } \sigma_z^{(i)}  + \sum\limits_{i < j,=1}^4 {\frac{\pi}{2}} J_{ij}  \sigma_z^{(i)} \otimes \sigma_z^{(j)},
\end{align}
with the chemical frequency $\nu_i$ and the $J$-coupling  value $J_{ij}$ between the $i$-th and $j$-th carbons. More details about samples and experiments can be found in \cite{sm}. 

As shown in Fig. \ref{fig3}(a), our experiments contain the following steps. (i) \emph{Initialization}.  We initialize the four-qubit NMR system from the thermal equilibrium state to the pseudo-pure state (PPS), $\rho_{\text{pps}} \sim \ket{0}\bra{0}^{\otimes 4}$, using the spatial averaging technique \cite{lu2016tomography}. (ii) \emph{State engineering}. We prepare nine types of  local-unitary-equivalent states considering that four-qubit state can be always entangled in nine ways \cite{verstraete2002four, giordano2022reinforcement, vintskevich2023classification}. The preparation circuit typically includes single-qubit NOT gates, controlled-NOT gates, controlled-rotation gates, the Toffoli gate, and random local unitary operations, which are implemented using shaped radio-frequency pulses, optimized via a gradient ascent algorithm to ensure robustness against radio-frequency and static field inhomogeneities \cite{khaneja2005optimal}. Figure \ref{fig3}(a) illustrates the complete experimental protocol, using the preparation of an entangled state $\ket{\Psi_4} = (\ket{0000} + \ket{1111} + \ket{0101} + \ket{1010} + \ket{0110})/\sqrt{5}$ as an example. (iii) \emph{Measurement}. For each QOT measurement observable $\hat{g}_i$, we apply the readout operation $U_R$ to rotate the state and obtain its diagonal elements $\mu$ by measuring the expectation values of $\{I, Z\}^{\otimes 4}$. NMR enables the direct measurement of these expectation values from the experimental spectra of $^{13}$C, due to the feature of average measurement \cite{sm}. The expectation values of the observables $\hat{o}_i$, forming 2-RDMs, are computed by,
\begin{equation}
\label{Hamiltonian}
\left\langle \hat{o}_i \right\rangle =\sum_j^D(1-[ \hat{o}_i ,  \hat{g}_j ]_{\text{qwc}})\sum_{l=1}^{2^N} \mu^{(j)}_l \cdot \bra{l} U^j_R \hat{o}_i U^{j\dag}_R\ket{l},
\end{equation}
Let $\ket{l}$ represent the $l$-th computational basis. Here, we reconstruct the 2-RDMs using both the original QOT with $D=15$ observables and the optimal QOT with $D=9$ observables.

Quantum state classification, particularly for multi-qubit systems, is often challenging due to the complexity of full QST. Since QOT efficiently reconstructs RDMs and the multiparticle entanglement of RDMs reveals entanglement patterns, a natural idea is to use machine learning to classify states directly from QOT measurements. Our numerical and experimental results latter validate this intuition. As a proof-of-demonstration, we create a neural network model structured as input-(256-512-128)-output, in which the input consists of QOT measurement results and the output $P$ denotes the state classification value. Figure \ref{fig3}(c) illustrates our quantum state classifier.  We define the loss function by summing over the distance between the true and predicted $P$. Here, we first create 100, 000 local-unitary-equivalent four-qubit states $\ket{\psi'_4}=\otimes_{i=1}^4U^{(i)}_L(\theta_i)\ket{\psi_4}$ for each class, where $U^{(i)}_L(\theta_i)$ is a random local unitary and $\ket{\psi_4}$ belongs to one of nine classes of states  \cite{verstraete2002four}. We next numerically simulate noisy QOT measurements as input data. Then, the model is trained using gradient descent and the Adam optimizer. Once trained, the model classifies the experimentally prepared states based on QOT measurements, without the need for full QST.

\textit{Results.—} First, the original QOT requires approximately 67\% more experimental time than our approach when reconstructing 2-RDMs for four-qubit states, supporting the feasibility and advantage of our QOT method. Figure \ref{fig3}(b) displays the partial RDMs and their corresponding concurrence for the experimentally prepared state $\ket{\Psi_4}$ \cite{wootters2001entanglement}. The average fidelity with the expected RDMs is about 99.11\% for the original QOT and 99.10\% for the optimal QOT. The concurrence pattern can be revealed from the reconstructed RDMs. The concurrence is nearly zero when any two qubits are traced out, except when qubits 1 and 3 (2 and 4) are traced out. This behavior contrasts with the typical four-qubit GHZ  state, where all concurrences are zero, and the four-qubit W state, which exhibits a concurrence of 0.5 regardless of which two qubits are traced out. 

Second, Figure \ref{fig3}(d) shows the classification results. Here, we evaluate the performance by inputting 20, 000 simulation testdata for each class. Additionally, we experimentally prepare five random local-unitary-equivalent states for each class, in total 45 four-qubit states, and implement our QOT scheme. The QOT measurement data are not only used to reconstruct RDMs \cite{sm}, but can also be input into the neural network for classification. The results indicate classification accuracies of 95.9\% for the simulation data and 95.6\% for the experimental data, confirming the potential application of QOT in quantum state classification. This model allows us to directly reveal the entanglement pattern of the experimentally prepared states. For example, states in Class 2 exhibit non-zero concurrence for every pair of qubits, whereas states in Class 7 show non-zero concurrence only between qubits 2 and 3 (or 4).

Moreover, we demonstrate the adaptability of our method using noisy measurement data from superconducting quantum processors, highlighting the advantage in the number of measurement samples. The data includes sampling measurement results of different parallel observables required by both the original QOT and our QOT method for six- and nine-qubit experimentally prepared W states \cite{hu2024experimental}. Similarly, we compute the expectation values of the observables $\hat{o}_i$ that form each 2-RDM. Based on the Chernoff-Hoeffding inequality  \cite{phillips2012chernoff}, the norm distance between the reconstructed RDMs $\rho_k'$ and actual $\rho_k$ under $M$  samples can be written as,
\begin{align}
||\rho_k'-\rho_k|| =\frac{d}{\sqrt{M}}, ~\text{with}~d^2=\sum_{i=1}^{4^k} \frac{1-\left\langle \hat{o}_i \right \rangle^2}{\Lambda[\hat{o}_i]}.
\label{define_d}
\end{align}
$\Lambda[\hat{o}_i]$ is the probability of measuring the observable $\hat{o}_i$ in implementing QOT scheme. $d$ is referred to as the one-shot norm distance, which depends on the specific QOT scheme. The required  $M$ is proportional to $d^{2}$ when achieving the same norm distance. 

\begin{figure}
 \centering
  \includegraphics[width=0.49\textwidth]{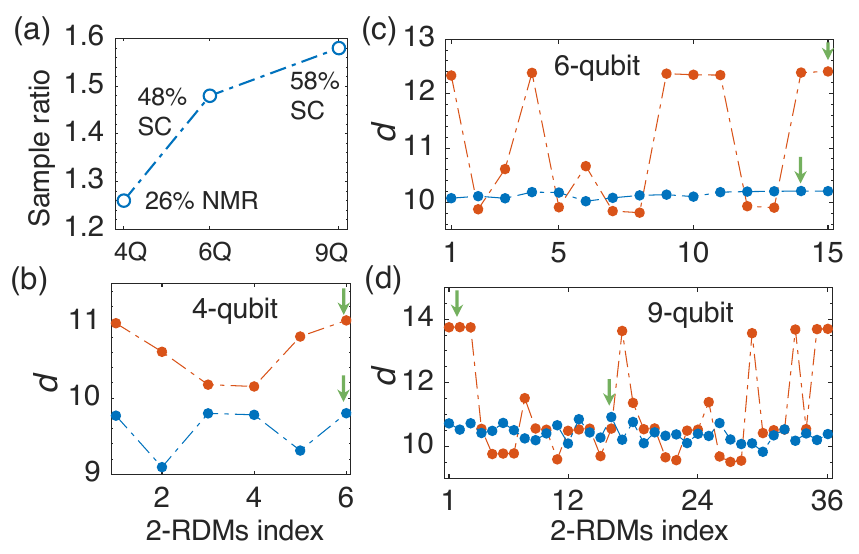}
  \caption{(a) The sample requirement ratio between the original QOT and our QOT schemes for reconstructing 2-RDMs of the four-, six-, and nine-qubit experimentally prepared states. (b) The $d$ value for each 2-RDM using the original QOT (red lines) and our QOT (blue lines) schemes on a four-qubit NMR platform. (c-d) The corresponding $d$ values for all 2-RDMs of six- and nine-qubit prepared W states on superconducting (SC) platforms. The 2-RDMs with worst norm distance are indicated using the green arrows. } 
  \label{fig4}
\end{figure}

Figure \ref{fig4}(b-d) presents the $d$ values for all 2-RDMs of four-, six-, and nine-qubit experimentally prepared states, utilizing both the original QOT and our QOT methods. While the mean $d$  for our method is slightly better to that of the original QOT, our aim is to accurately reconstruct all RDMs, rather than solely concentrating on the average value. Consequently, we focus on the worst one-shot norm distance, where the original QOT shows a larger $d$ compared to our method. Additionally, we compare the sample requirements between the two schemes to achieve the same norm distance, with the ratio displayed in Fig. \ref{fig4}(a). The results indicate that the original QOT needs 26\%, 48\%, and 58\% more samples than our method for four-, six-, and nine-qubit states, with this discrepancy becoming more obvious as the system size increases.

\textit{Conclusions.—}  We propose a more efficient QOT method that not only reduces the number of measurement settings and samples required but also enhances flexibility and scalability in reconstructing RDMs compared to the original QOT.  Additionally, we experimentally verify its feasibility, advantages, and applications across different quantum processors. Here, we assume an equal number of samples is measured for each measurement setting. Indeed, the number of local observables that can be measured varies across different measurement settings. Further improvements can be achieved by allocating more samples to those measurement settings that capture a greater number of local observables. As an efficient method for partial tomography, our method is well-suited for experimental quantum state characterization, determining quantum state properties such as expectation values of target observables and $k$-local Hamiltonians, and may even extend to tasks beyond quantum information science in the future  \cite{chang2020hybrid}.

%\begin{acknowledgments}
\textit{Note.—} While conducting this project, we found a related work posted on arXiv that also proposes an optimal QOT method for reconstructing RDMs \cite{hansenne2024optimal}. Their work appears similar to ours,  as it utilizes graph theory and demonstrates feasibility with a six-photon Dicke state. Our work employs the minimum clique cover to tackle this problem and validates its feasibility and applications using four-, six-, nine-qubit experimentally prepared states. Two years ago, we filed a Chinese patent for our method (No: \href{http://epub.cnipa.gov.cn/patent/CN114579926B}{2022101690085}).

\textit{Acknowledgments.—}This work is supported by the National Natural Science Foundation of China (12275117), Guangdong Basic and Applied Basic Research Foundation (2022B1515020074), Guangdong Provincial Key Laboratory (2019B121203002), Shenzhen Science and Technology Program (RCYX20200714114522109 and KQTD20200820113010023), and Center for Computational Science and Engineering at Southern University of Science and Technology. 
%\end{acknowledgments}

%\bibliography{OQOT.bib}

\begin{thebibliography}{53}%
\makeatletter
\providecommand \@ifxundefined [1]{%
 \@ifx{#1\undefined}
}%
\providecommand \@ifnum [1]{%
 \ifnum #1\expandafter \@firstoftwo
 \else \expandafter \@secondoftwo
 \fi
}%
\providecommand \@ifx [1]{%
 \ifx #1\expandafter \@firstoftwo
 \else \expandafter \@secondoftwo
 \fi
}%
\providecommand \natexlab [1]{#1}%
\providecommand \enquote  [1]{``#1''}%
\providecommand \bibnamefont  [1]{#1}%
\providecommand \bibfnamefont [1]{#1}%
\providecommand \citenamefont [1]{#1}%
\providecommand \href@noop [0]{\@secondoftwo}%
\providecommand \href [0]{\begingroup \@sanitize@url \@href}%
\providecommand \@href[1]{\@@startlink{#1}\@@href}%
\providecommand \@@href[1]{\endgroup#1\@@endlink}%
\providecommand \@sanitize@url [0]{\catcode `\\12\catcode `\$12\catcode
  `\&12\catcode `\#12\catcode `\^12\catcode `\_12\catcode `\%12\relax}%
\providecommand \@@startlink[1]{}%
\providecommand \@@endlink[0]{}%
\providecommand \url  [0]{\begingroup\@sanitize@url \@url }%
\providecommand \@url [1]{\endgroup\@href {#1}{\urlprefix }}%
\providecommand \urlprefix  [0]{URL }%
\providecommand \Eprint [0]{\href }%
\providecommand \doibase [0]{http://dx.doi.org/}%
\providecommand \selectlanguage [0]{\@gobble}%
\providecommand \bibinfo  [0]{\@secondoftwo}%
\providecommand \bibfield  [0]{\@secondoftwo}%
\providecommand \translation [1]{[#1]}%
\providecommand \BibitemOpen [0]{}%
\providecommand \bibitemStop [0]{}%
\providecommand \bibitemNoStop [0]{.\EOS\space}%
\providecommand \EOS [0]{\spacefactor3000\relax}%
\providecommand \BibitemShut  [1]{\csname bibitem#1\endcsname}%
\let\auto@bib@innerbib\@empty
%</preamble>
\bibitem [{\citenamefont {James}\ \emph {et~al.}(2001)\citenamefont {James},
  \citenamefont {Kwiat}, \citenamefont {Munro},\ and\ \citenamefont
  {White}}]{james2001measurement}%
  \BibitemOpen
  \bibfield  {author} {\bibinfo {author} {\bibfnamefont {D.~F.}\ \bibnamefont
  {James}}, \bibinfo {author} {\bibfnamefont {P.~G.}\ \bibnamefont {Kwiat}},
  \bibinfo {author} {\bibfnamefont {W.~J.}\ \bibnamefont {Munro}}, \ and\
  \bibinfo {author} {\bibfnamefont {A.~G.}\ \bibnamefont {White}},\ }\href
  {https://journals.aps.org/pra/abstract/10.1103/PhysRevA.64.052312} {\bibfield
   {journal} {\bibinfo  {journal} {Phys. Rev. A.}\ }\textbf {\bibinfo {volume}
  {64}},\ \bibinfo {pages} {052312} (\bibinfo {year} {2001})}\BibitemShut
  {NoStop}%
\bibitem [{\citenamefont {Merkel}\ \emph {et~al.}(2013)\citenamefont {Merkel},
  \citenamefont {Gambetta}, \citenamefont {Smolin}, \citenamefont {Poletto},
  \citenamefont {C{\'o}rcoles}, \citenamefont {Johnson}, \citenamefont {Ryan},\
  and\ \citenamefont {Steffen}}]{merkel2013self}%
  \BibitemOpen
  \bibfield  {author} {\bibinfo {author} {\bibfnamefont {S.~T.}\ \bibnamefont
  {Merkel}}, \bibinfo {author} {\bibfnamefont {J.~M.}\ \bibnamefont
  {Gambetta}}, \bibinfo {author} {\bibfnamefont {J.~A.}\ \bibnamefont
  {Smolin}}, \bibinfo {author} {\bibfnamefont {S.}~\bibnamefont {Poletto}},
  \bibinfo {author} {\bibfnamefont {A.~D.}\ \bibnamefont {C{\'o}rcoles}},
  \bibinfo {author} {\bibfnamefont {B.~R.}\ \bibnamefont {Johnson}}, \bibinfo
  {author} {\bibfnamefont {C.~A.}\ \bibnamefont {Ryan}}, \ and\ \bibinfo
  {author} {\bibfnamefont {M.}~\bibnamefont {Steffen}},\ }\href
  {https://journals.aps.org/pra/abstract/10.1103/PhysRevA.87.062119} {\bibfield
   {journal} {\bibinfo  {journal} {Phys. Rev. A.}\ }\textbf {\bibinfo {volume}
  {87}},\ \bibinfo {pages} {062119} (\bibinfo {year} {2013})}\BibitemShut
  {NoStop}%
\bibitem [{\citenamefont {Eisert}\ \emph {et~al.}(2020)\citenamefont {Eisert},
  \citenamefont {Hangleiter}, \citenamefont {Walk}, \citenamefont {Roth},
  \citenamefont {Markham}, \citenamefont {Parekh}, \citenamefont {Chabaud},\
  and\ \citenamefont {Kashefi}}]{eisert2020quantum}%
  \BibitemOpen
  \bibfield  {author} {\bibinfo {author} {\bibfnamefont {J.}~\bibnamefont
  {Eisert}}, \bibinfo {author} {\bibfnamefont {D.}~\bibnamefont {Hangleiter}},
  \bibinfo {author} {\bibfnamefont {N.}~\bibnamefont {Walk}}, \bibinfo {author}
  {\bibfnamefont {I.}~\bibnamefont {Roth}}, \bibinfo {author} {\bibfnamefont
  {D.}~\bibnamefont {Markham}}, \bibinfo {author} {\bibfnamefont
  {R.}~\bibnamefont {Parekh}}, \bibinfo {author} {\bibfnamefont
  {U.}~\bibnamefont {Chabaud}}, \ and\ \bibinfo {author} {\bibfnamefont
  {E.}~\bibnamefont {Kashefi}},\ }\href
  {https://www.nature.com/articles/s42254-020-0186-4} {\bibfield  {journal}
  {\bibinfo  {journal} {Nat. Rev. Phys.}\ }\textbf {\bibinfo {volume} {2}},\
  \bibinfo {pages} {382} (\bibinfo {year} {2020})}\BibitemShut {NoStop}%
\bibitem [{\citenamefont {Cramer}\ \emph {et~al.}(2010)\citenamefont {Cramer},
  \citenamefont {Plenio}, \citenamefont {Flammia}, \citenamefont {Somma},
  \citenamefont {Gross}, \citenamefont {Bartlett}, \citenamefont
  {Landon-Cardinal}, \citenamefont {Poulin},\ and\ \citenamefont
  {Liu}}]{cramer2010efficient}%
  \BibitemOpen
  \bibfield  {author} {\bibinfo {author} {\bibfnamefont {M.}~\bibnamefont
  {Cramer}}, \bibinfo {author} {\bibfnamefont {M.~B.}\ \bibnamefont {Plenio}},
  \bibinfo {author} {\bibfnamefont {S.~T.}\ \bibnamefont {Flammia}}, \bibinfo
  {author} {\bibfnamefont {R.}~\bibnamefont {Somma}}, \bibinfo {author}
  {\bibfnamefont {D.}~\bibnamefont {Gross}}, \bibinfo {author} {\bibfnamefont
  {S.~D.}\ \bibnamefont {Bartlett}}, \bibinfo {author} {\bibfnamefont
  {O.}~\bibnamefont {Landon-Cardinal}}, \bibinfo {author} {\bibfnamefont
  {D.}~\bibnamefont {Poulin}}, \ and\ \bibinfo {author} {\bibfnamefont {Y.-K.}\
  \bibnamefont {Liu}},\ }\href {https://www.nature.com/articles/ncomms1147}
  {\bibfield  {journal} {\bibinfo  {journal} {Nat. Commun.}\ }\textbf {\bibinfo
  {volume} {1}},\ \bibinfo {pages} {149} (\bibinfo {year} {2010})}\BibitemShut
  {NoStop}%
\bibitem [{\citenamefont {Flammia}\ \emph {et~al.}(2012)\citenamefont
  {Flammia}, \citenamefont {Gross}, \citenamefont {Liu},\ and\ \citenamefont
  {Eisert}}]{flammia2012quantum}%
  \BibitemOpen
  \bibfield  {author} {\bibinfo {author} {\bibfnamefont {S.~T.}\ \bibnamefont
  {Flammia}}, \bibinfo {author} {\bibfnamefont {D.}~\bibnamefont {Gross}},
  \bibinfo {author} {\bibfnamefont {Y.-K.}\ \bibnamefont {Liu}}, \ and\
  \bibinfo {author} {\bibfnamefont {J.}~\bibnamefont {Eisert}},\ }\href
  {https://iopscience.iop.org/article/10.1088/1367-2630/14/9/095022/meta}
  {\bibfield  {journal} {\bibinfo  {journal} {New J. Phys.}\ }\textbf {\bibinfo
  {volume} {14}},\ \bibinfo {pages} {095022} (\bibinfo {year}
  {2012})}\BibitemShut {NoStop}%
\bibitem [{\citenamefont {Lanyon}\ \emph {et~al.}(2017)\citenamefont {Lanyon},
  \citenamefont {Maier}, \citenamefont {Holz{\"a}pfel}, \citenamefont
  {Baumgratz}, \citenamefont {Hempel}, \citenamefont {Jurcevic}, \citenamefont
  {Dhand}, \citenamefont {Buyskikh}, \citenamefont {Daley}, \citenamefont
  {Cramer} \emph {et~al.}}]{lanyon2017efficient}%
  \BibitemOpen
  \bibfield  {author} {\bibinfo {author} {\bibfnamefont {B.~P.}\ \bibnamefont
  {Lanyon}}, \bibinfo {author} {\bibfnamefont {C.}~\bibnamefont {Maier}},
  \bibinfo {author} {\bibfnamefont {M.}~\bibnamefont {Holz{\"a}pfel}}, \bibinfo
  {author} {\bibfnamefont {T.}~\bibnamefont {Baumgratz}}, \bibinfo {author}
  {\bibfnamefont {C.}~\bibnamefont {Hempel}}, \bibinfo {author} {\bibfnamefont
  {P.}~\bibnamefont {Jurcevic}}, \bibinfo {author} {\bibfnamefont
  {I.}~\bibnamefont {Dhand}}, \bibinfo {author} {\bibfnamefont
  {A.}~\bibnamefont {Buyskikh}}, \bibinfo {author} {\bibfnamefont {A.~J.}\
  \bibnamefont {Daley}}, \bibinfo {author} {\bibfnamefont {M.}~\bibnamefont
  {Cramer}},  \emph {et~al.},\ }\href
  {https://www.nature.com/articles/nphys4244} {\bibfield  {journal} {\bibinfo
  {journal} {Nat. Phys.}\ }\textbf {\bibinfo {volume} {13}},\ \bibinfo {pages}
  {1158} (\bibinfo {year} {2017})}\BibitemShut {NoStop}%
\bibitem [{\citenamefont {Huang}\ \emph {et~al.}(2020)\citenamefont {Huang},
  \citenamefont {Kueng},\ and\ \citenamefont {Preskill}}]{huang2020predicting}%
  \BibitemOpen
  \bibfield  {author} {\bibinfo {author} {\bibfnamefont {H.-Y.}\ \bibnamefont
  {Huang}}, \bibinfo {author} {\bibfnamefont {R.}~\bibnamefont {Kueng}}, \ and\
  \bibinfo {author} {\bibfnamefont {J.}~\bibnamefont {Preskill}},\ }\href
  {https://www.nature.com/articles/s41567-020-0932-7} {\bibfield  {journal}
  {\bibinfo  {journal} {Nat. Phys.}\ }\textbf {\bibinfo {volume} {16}},\
  \bibinfo {pages} {1050} (\bibinfo {year} {2020})}\BibitemShut {NoStop}%
\bibitem [{\citenamefont {Song}\ \emph {et~al.}(2017)\citenamefont {Song},
  \citenamefont {Xu}, \citenamefont {Liu}, \citenamefont {Yang}, \citenamefont
  {Zheng}, \citenamefont {Deng}, \citenamefont {Xie}, \citenamefont {Huang},
  \citenamefont {Guo}, \citenamefont {Zhang} \emph {et~al.}}]{song201710}%
  \BibitemOpen
  \bibfield  {author} {\bibinfo {author} {\bibfnamefont {C.}~\bibnamefont
  {Song}}, \bibinfo {author} {\bibfnamefont {K.}~\bibnamefont {Xu}}, \bibinfo
  {author} {\bibfnamefont {W.}~\bibnamefont {Liu}}, \bibinfo {author}
  {\bibfnamefont {C.-p.}\ \bibnamefont {Yang}}, \bibinfo {author}
  {\bibfnamefont {S.-B.}\ \bibnamefont {Zheng}}, \bibinfo {author}
  {\bibfnamefont {H.}~\bibnamefont {Deng}}, \bibinfo {author} {\bibfnamefont
  {Q.}~\bibnamefont {Xie}}, \bibinfo {author} {\bibfnamefont {K.}~\bibnamefont
  {Huang}}, \bibinfo {author} {\bibfnamefont {Q.}~\bibnamefont {Guo}}, \bibinfo
  {author} {\bibfnamefont {L.}~\bibnamefont {Zhang}},  \emph {et~al.},\ }\href
  {https://journals.aps.org/prl/abstract/10.1103/PhysRevLett.119.180511}
  {\bibfield  {journal} {\bibinfo  {journal} {Phys. Rev. Lett.}\ }\textbf
  {\bibinfo {volume} {119}},\ \bibinfo {pages} {180511} (\bibinfo {year}
  {2017})}\BibitemShut {NoStop}%
\bibitem [{\citenamefont {Di{\'o}si}(2004)}]{diosi2004three}%
  \BibitemOpen
  \bibfield  {author} {\bibinfo {author} {\bibfnamefont {L.}~\bibnamefont
  {Di{\'o}si}},\ }\href
  {https://journals.aps.org/pra/abstract/10.1103/PhysRevA.70.010302} {\bibfield
   {journal} {\bibinfo  {journal} {Phys. Rev. A.}\ }\textbf {\bibinfo {volume}
  {70}},\ \bibinfo {pages} {010302} (\bibinfo {year} {2004})}\BibitemShut
  {NoStop}%
\bibitem [{\citenamefont {Chen}\ \emph {et~al.}(2013)\citenamefont {Chen},
  \citenamefont {Dawkins}, \citenamefont {Ji}, \citenamefont {Johnston},
  \citenamefont {Kribs}, \citenamefont {Shultz},\ and\ \citenamefont
  {Zeng}}]{chen2013uniqueness}%
  \BibitemOpen
  \bibfield  {author} {\bibinfo {author} {\bibfnamefont {J.}~\bibnamefont
  {Chen}}, \bibinfo {author} {\bibfnamefont {H.}~\bibnamefont {Dawkins}},
  \bibinfo {author} {\bibfnamefont {Z.}~\bibnamefont {Ji}}, \bibinfo {author}
  {\bibfnamefont {N.}~\bibnamefont {Johnston}}, \bibinfo {author}
  {\bibfnamefont {D.}~\bibnamefont {Kribs}}, \bibinfo {author} {\bibfnamefont
  {F.}~\bibnamefont {Shultz}}, \ and\ \bibinfo {author} {\bibfnamefont
  {B.}~\bibnamefont {Zeng}},\ }\href
  {https://journals.aps.org/pra/abstract/10.1103/PhysRevA.88.012109} {\bibfield
   {journal} {\bibinfo  {journal} {Phys. Rev. A.}\ }\textbf {\bibinfo {volume}
  {88}},\ \bibinfo {pages} {012109} (\bibinfo {year} {2013})}\BibitemShut
  {NoStop}%
\bibitem [{\citenamefont {Xin}\ \emph {et~al.}(2017)\citenamefont {Xin},
  \citenamefont {Lu}, \citenamefont {Klassen}, \citenamefont {Yu},
  \citenamefont {Ji}, \citenamefont {Chen}, \citenamefont {Ma}, \citenamefont
  {Long}, \citenamefont {Zeng},\ and\ \citenamefont
  {Laflamme}}]{xin2017quantum}%
  \BibitemOpen
  \bibfield  {author} {\bibinfo {author} {\bibfnamefont {T.}~\bibnamefont
  {Xin}}, \bibinfo {author} {\bibfnamefont {D.}~\bibnamefont {Lu}}, \bibinfo
  {author} {\bibfnamefont {J.}~\bibnamefont {Klassen}}, \bibinfo {author}
  {\bibfnamefont {N.}~\bibnamefont {Yu}}, \bibinfo {author} {\bibfnamefont
  {Z.}~\bibnamefont {Ji}}, \bibinfo {author} {\bibfnamefont {J.}~\bibnamefont
  {Chen}}, \bibinfo {author} {\bibfnamefont {X.}~\bibnamefont {Ma}}, \bibinfo
  {author} {\bibfnamefont {G.}~\bibnamefont {Long}}, \bibinfo {author}
  {\bibfnamefont {B.}~\bibnamefont {Zeng}}, \ and\ \bibinfo {author}
  {\bibfnamefont {R.}~\bibnamefont {Laflamme}},\ }\href
  {https://journals.aps.org/prl/abstract/10.1103/PhysRevLett.118.020401}
  {\bibfield  {journal} {\bibinfo  {journal} {Phys. Rev. Lett.}\ }\textbf
  {\bibinfo {volume} {118}},\ \bibinfo {pages} {020401} (\bibinfo {year}
  {2017})}\BibitemShut {NoStop}%
\bibitem [{\citenamefont {Guo}\ and\ \citenamefont
  {Yang}(2024)}]{guo2024quantum}%
  \BibitemOpen
  \bibfield  {author} {\bibinfo {author} {\bibfnamefont {Y.}~\bibnamefont
  {Guo}}\ and\ \bibinfo {author} {\bibfnamefont {S.}~\bibnamefont {Yang}},\
  }\href {https://www.nature.com/articles/s42005-024-01813-4} {\bibfield
  {journal} {\bibinfo  {journal} {Commun. Phys.}\ }\textbf {\bibinfo {volume}
  {7}},\ \bibinfo {pages} {322} (\bibinfo {year} {2024})}\BibitemShut {NoStop}%
\bibitem [{\citenamefont {Flammia}\ and\ \citenamefont
  {Liu}(2011)}]{flammia2011direct}%
  \BibitemOpen
  \bibfield  {author} {\bibinfo {author} {\bibfnamefont {S.~T.}\ \bibnamefont
  {Flammia}}\ and\ \bibinfo {author} {\bibfnamefont {Y.-K.}\ \bibnamefont
  {Liu}},\ }\href
  {https://journals.aps.org/prl/abstract/10.1103/PhysRevLett.106.230501}
  {\bibfield  {journal} {\bibinfo  {journal} {Phys. Rev. Lett.}\ }\textbf
  {\bibinfo {volume} {106}},\ \bibinfo {pages} {230501} (\bibinfo {year}
  {2011})}\BibitemShut {NoStop}%
\bibitem [{\citenamefont {Cerezo}\ \emph {et~al.}(2020)\citenamefont {Cerezo},
  \citenamefont {Poremba}, \citenamefont {Cincio},\ and\ \citenamefont
  {Coles}}]{cerezo2020variational}%
  \BibitemOpen
  \bibfield  {author} {\bibinfo {author} {\bibfnamefont {M.}~\bibnamefont
  {Cerezo}}, \bibinfo {author} {\bibfnamefont {A.}~\bibnamefont {Poremba}},
  \bibinfo {author} {\bibfnamefont {L.}~\bibnamefont {Cincio}}, \ and\ \bibinfo
  {author} {\bibfnamefont {P.~J.}\ \bibnamefont {Coles}},\ }\href
  {https://quantum-journal.org/papers/q-2020-03-26-248/} {\bibfield  {journal}
  {\bibinfo  {journal} {Quantum}\ }\textbf {\bibinfo {volume} {4}},\ \bibinfo
  {pages} {248} (\bibinfo {year} {2020})}\BibitemShut {NoStop}%
\bibitem [{\citenamefont {Wu}\ \emph {et~al.}(2023)\citenamefont {Wu},
  \citenamefont {Zhu}, \citenamefont {Bai}, \citenamefont {Wang},\ and\
  \citenamefont {Chiribella}}]{wu2023quantum}%
  \BibitemOpen
  \bibfield  {author} {\bibinfo {author} {\bibfnamefont {Y.-D.}\ \bibnamefont
  {Wu}}, \bibinfo {author} {\bibfnamefont {Y.}~\bibnamefont {Zhu}}, \bibinfo
  {author} {\bibfnamefont {G.}~\bibnamefont {Bai}}, \bibinfo {author}
  {\bibfnamefont {Y.}~\bibnamefont {Wang}}, \ and\ \bibinfo {author}
  {\bibfnamefont {G.}~\bibnamefont {Chiribella}},\ }\href
  {https://journals.aps.org/prl/abstract/10.1103/PhysRevLett.130.210601}
  {\bibfield  {journal} {\bibinfo  {journal} {Phys. Rev. Lett.}\ }\textbf
  {\bibinfo {volume} {130}},\ \bibinfo {pages} {210601} (\bibinfo {year}
  {2023})}\BibitemShut {NoStop}%
\bibitem [{\citenamefont {Qin}\ \emph {et~al.}(2024)\citenamefont {Qin},
  \citenamefont {Che}, \citenamefont {Wei}, \citenamefont {Xu}, \citenamefont
  {Huang},\ and\ \citenamefont {Xin}}]{qin2024experimental}%
  \BibitemOpen
  \bibfield  {author} {\bibinfo {author} {\bibfnamefont {H.}~\bibnamefont
  {Qin}}, \bibinfo {author} {\bibfnamefont {L.}~\bibnamefont {Che}}, \bibinfo
  {author} {\bibfnamefont {C.}~\bibnamefont {Wei}}, \bibinfo {author}
  {\bibfnamefont {F.}~\bibnamefont {Xu}}, \bibinfo {author} {\bibfnamefont
  {Y.}~\bibnamefont {Huang}}, \ and\ \bibinfo {author} {\bibfnamefont
  {T.}~\bibnamefont {Xin}},\ }\href
  {https://journals.aps.org/prl/abstract/10.1103/PhysRevLett.132.190801}
  {\bibfield  {journal} {\bibinfo  {journal} {Phys. Rev. Lett.}\ }\textbf
  {\bibinfo {volume} {132}},\ \bibinfo {pages} {190801} (\bibinfo {year}
  {2024})}\BibitemShut {NoStop}%
\bibitem [{\citenamefont {Islam}\ \emph {et~al.}(2015)\citenamefont {Islam},
  \citenamefont {Ma}, \citenamefont {Preiss}, \citenamefont {Eric~Tai},
  \citenamefont {Lukin}, \citenamefont {Rispoli},\ and\ \citenamefont
  {Greiner}}]{islam2015measuring}%
  \BibitemOpen
  \bibfield  {author} {\bibinfo {author} {\bibfnamefont {R.}~\bibnamefont
  {Islam}}, \bibinfo {author} {\bibfnamefont {R.}~\bibnamefont {Ma}}, \bibinfo
  {author} {\bibfnamefont {P.~M.}\ \bibnamefont {Preiss}}, \bibinfo {author}
  {\bibfnamefont {M.}~\bibnamefont {Eric~Tai}}, \bibinfo {author}
  {\bibfnamefont {A.}~\bibnamefont {Lukin}}, \bibinfo {author} {\bibfnamefont
  {M.}~\bibnamefont {Rispoli}}, \ and\ \bibinfo {author} {\bibfnamefont
  {M.}~\bibnamefont {Greiner}},\ }\href
  {https://www.nature.com/articles/nature15750} {\bibfield  {journal} {\bibinfo
   {journal} {Nature}\ }\textbf {\bibinfo {volume} {528}},\ \bibinfo {pages}
  {77} (\bibinfo {year} {2015})}\BibitemShut {NoStop}%
\bibitem [{\citenamefont {Elben}\ \emph {et~al.}(2020)\citenamefont {Elben},
  \citenamefont {Kueng}, \citenamefont {Huang}, \citenamefont {van Bijnen},
  \citenamefont {Kokail}, \citenamefont {Dalmonte}, \citenamefont {Calabrese},
  \citenamefont {Kraus}, \citenamefont {Preskill}, \citenamefont {Zoller} \emph
  {et~al.}}]{elben2020mixed}%
  \BibitemOpen
  \bibfield  {author} {\bibinfo {author} {\bibfnamefont {A.}~\bibnamefont
  {Elben}}, \bibinfo {author} {\bibfnamefont {R.}~\bibnamefont {Kueng}},
  \bibinfo {author} {\bibfnamefont {H.-Y.}\ \bibnamefont {Huang}}, \bibinfo
  {author} {\bibfnamefont {R.}~\bibnamefont {van Bijnen}}, \bibinfo {author}
  {\bibfnamefont {C.}~\bibnamefont {Kokail}}, \bibinfo {author} {\bibfnamefont
  {M.}~\bibnamefont {Dalmonte}}, \bibinfo {author} {\bibfnamefont
  {P.}~\bibnamefont {Calabrese}}, \bibinfo {author} {\bibfnamefont
  {B.}~\bibnamefont {Kraus}}, \bibinfo {author} {\bibfnamefont
  {J.}~\bibnamefont {Preskill}}, \bibinfo {author} {\bibfnamefont
  {P.}~\bibnamefont {Zoller}},  \emph {et~al.},\ }\href
  {https://journals.aps.org/prl/abstract/10.1103/PhysRevLett.125.200501}
  {\bibfield  {journal} {\bibinfo  {journal} {Phys. Rev. Lett.}\ }\textbf
  {\bibinfo {volume} {125}},\ \bibinfo {pages} {200501} (\bibinfo {year}
  {2020})}\BibitemShut {NoStop}%
\bibitem [{\citenamefont {Huang}\ \emph {et~al.}(2022)\citenamefont {Huang},
  \citenamefont {Che}, \citenamefont {Wei}, \citenamefont {Xu}, \citenamefont
  {Nie}, \citenamefont {Li}, \citenamefont {Lu},\ and\ \citenamefont
  {Xin}}]{huang2022measuring}%
  \BibitemOpen
  \bibfield  {author} {\bibinfo {author} {\bibfnamefont {Y.}~\bibnamefont
  {Huang}}, \bibinfo {author} {\bibfnamefont {L.}~\bibnamefont {Che}}, \bibinfo
  {author} {\bibfnamefont {C.}~\bibnamefont {Wei}}, \bibinfo {author}
  {\bibfnamefont {F.}~\bibnamefont {Xu}}, \bibinfo {author} {\bibfnamefont
  {X.}~\bibnamefont {Nie}}, \bibinfo {author} {\bibfnamefont {J.}~\bibnamefont
  {Li}}, \bibinfo {author} {\bibfnamefont {D.}~\bibnamefont {Lu}}, \ and\
  \bibinfo {author} {\bibfnamefont {T.}~\bibnamefont {Xin}},\ }\href
  {https://arxiv.org/abs/2209.08501} {\bibfield  {journal} {\bibinfo  {journal}
  {arXiv:2209.08501}\ } (\bibinfo {year} {2022})}\BibitemShut {NoStop}%
\bibitem [{\citenamefont {Wu}\ \emph {et~al.}(2024)\citenamefont {Wu},
  \citenamefont {Zhu}, \citenamefont {Wang},\ and\ \citenamefont
  {Chiribella}}]{wu2023learning}%
  \BibitemOpen
  \bibfield  {author} {\bibinfo {author} {\bibfnamefont {Y.-D.}\ \bibnamefont
  {Wu}}, \bibinfo {author} {\bibfnamefont {Y.}~\bibnamefont {Zhu}}, \bibinfo
  {author} {\bibfnamefont {Y.}~\bibnamefont {Wang}}, \ and\ \bibinfo {author}
  {\bibfnamefont {G.}~\bibnamefont {Chiribella}},\ }\href
  {https://www.nature.com/articles/s41467-024-53101-y} {\bibfield  {journal}
  {\bibinfo  {journal} {Nat. Commun.}\ }\textbf {\bibinfo {volume} {15}},\
  \bibinfo {pages} {8796} (\bibinfo {year} {2024})}\BibitemShut {NoStop}%
\bibitem [{\citenamefont {Wang}\ \emph {et~al.}(2015)\citenamefont {Wang},
  \citenamefont {Deng},\ and\ \citenamefont {Duan}}]{wang2015hamiltonian}%
  \BibitemOpen
  \bibfield  {author} {\bibinfo {author} {\bibfnamefont {S.-T.}\ \bibnamefont
  {Wang}}, \bibinfo {author} {\bibfnamefont {D.-L.}\ \bibnamefont {Deng}}, \
  and\ \bibinfo {author} {\bibfnamefont {L.-M.}\ \bibnamefont {Duan}},\ }\href
  {https://iopscience.iop.org/article/10.1088/1367-2630/17/9/093017/meta}
  {\bibfield  {journal} {\bibinfo  {journal} {New J. Phys.}\ }\textbf {\bibinfo
  {volume} {17}},\ \bibinfo {pages} {093017} (\bibinfo {year}
  {2015})}\BibitemShut {NoStop}%
\bibitem [{\citenamefont {Xin}\ \emph {et~al.}(2019)\citenamefont {Xin},
  \citenamefont {Lu}, \citenamefont {Cao}, \citenamefont {Anikeeva},
  \citenamefont {Lu}, \citenamefont {Li}, \citenamefont {Long},\ and\
  \citenamefont {Zeng}}]{xin2019local}%
  \BibitemOpen
  \bibfield  {author} {\bibinfo {author} {\bibfnamefont {T.}~\bibnamefont
  {Xin}}, \bibinfo {author} {\bibfnamefont {S.}~\bibnamefont {Lu}}, \bibinfo
  {author} {\bibfnamefont {N.}~\bibnamefont {Cao}}, \bibinfo {author}
  {\bibfnamefont {G.}~\bibnamefont {Anikeeva}}, \bibinfo {author}
  {\bibfnamefont {D.}~\bibnamefont {Lu}}, \bibinfo {author} {\bibfnamefont
  {J.}~\bibnamefont {Li}}, \bibinfo {author} {\bibfnamefont {G.}~\bibnamefont
  {Long}}, \ and\ \bibinfo {author} {\bibfnamefont {B.}~\bibnamefont {Zeng}},\
  }\href {https://www.nature.com/articles/s41534-019-0222-3} {\bibfield
  {journal} {\bibinfo  {journal} {npj Quantum Inf.}\ }\textbf {\bibinfo
  {volume} {5}},\ \bibinfo {pages} {109} (\bibinfo {year} {2019})}\BibitemShut
  {NoStop}%
\bibitem [{\citenamefont {Che}\ \emph {et~al.}(2021)\citenamefont {Che},
  \citenamefont {Wei}, \citenamefont {Huang}, \citenamefont {Zhao},
  \citenamefont {Xue}, \citenamefont {Nie}, \citenamefont {Li}, \citenamefont
  {Lu},\ and\ \citenamefont {Xin}}]{che2021learning}%
  \BibitemOpen
  \bibfield  {author} {\bibinfo {author} {\bibfnamefont {L.}~\bibnamefont
  {Che}}, \bibinfo {author} {\bibfnamefont {C.}~\bibnamefont {Wei}}, \bibinfo
  {author} {\bibfnamefont {Y.}~\bibnamefont {Huang}}, \bibinfo {author}
  {\bibfnamefont {D.}~\bibnamefont {Zhao}}, \bibinfo {author} {\bibfnamefont
  {S.}~\bibnamefont {Xue}}, \bibinfo {author} {\bibfnamefont {X.}~\bibnamefont
  {Nie}}, \bibinfo {author} {\bibfnamefont {J.}~\bibnamefont {Li}}, \bibinfo
  {author} {\bibfnamefont {D.}~\bibnamefont {Lu}}, \ and\ \bibinfo {author}
  {\bibfnamefont {T.}~\bibnamefont {Xin}},\ }\href
  {https://journals.aps.org/prresearch/abstract/10.1103/PhysRevResearch.3.023246}
  {\bibfield  {journal} {\bibinfo  {journal} {Phys. Rev. Res.}\ }\textbf
  {\bibinfo {volume} {3}},\ \bibinfo {pages} {023246} (\bibinfo {year}
  {2021})}\BibitemShut {NoStop}%
\bibitem [{\citenamefont {Peruzzo}\ \emph {et~al.}(2014)\citenamefont
  {Peruzzo}, \citenamefont {McClean}, \citenamefont {Shadbolt}, \citenamefont
  {Yung}, \citenamefont {Zhou}, \citenamefont {Love}, \citenamefont
  {Aspuru-Guzik},\ and\ \citenamefont {O’brien}}]{peruzzo2014variational}%
  \BibitemOpen
  \bibfield  {author} {\bibinfo {author} {\bibfnamefont {A.}~\bibnamefont
  {Peruzzo}}, \bibinfo {author} {\bibfnamefont {J.}~\bibnamefont {McClean}},
  \bibinfo {author} {\bibfnamefont {P.}~\bibnamefont {Shadbolt}}, \bibinfo
  {author} {\bibfnamefont {M.-H.}\ \bibnamefont {Yung}}, \bibinfo {author}
  {\bibfnamefont {X.-Q.}\ \bibnamefont {Zhou}}, \bibinfo {author}
  {\bibfnamefont {P.~J.}\ \bibnamefont {Love}}, \bibinfo {author}
  {\bibfnamefont {A.}~\bibnamefont {Aspuru-Guzik}}, \ and\ \bibinfo {author}
  {\bibfnamefont {J.~L.}\ \bibnamefont {O’brien}},\ }\href
  {https://www.nature.com/articles/ncomms5213} {\bibfield  {journal} {\bibinfo
  {journal} {Nat. Commun.}\ }\textbf {\bibinfo {volume} {5}},\ \bibinfo {pages}
  {4213} (\bibinfo {year} {2014})}\BibitemShut {NoStop}%
\bibitem [{\citenamefont {Verteletskyi}\ \emph {et~al.}(2020)\citenamefont
  {Verteletskyi}, \citenamefont {Yen},\ and\ \citenamefont
  {Izmaylov}}]{verteletskyi2020measurement}%
  \BibitemOpen
  \bibfield  {author} {\bibinfo {author} {\bibfnamefont {V.}~\bibnamefont
  {Verteletskyi}}, \bibinfo {author} {\bibfnamefont {T.-C.}\ \bibnamefont
  {Yen}}, \ and\ \bibinfo {author} {\bibfnamefont {A.~F.}\ \bibnamefont
  {Izmaylov}},\ }\href
  {https://pubs.aip.org/aip/jcp/article/152/12/124114/954934} {\bibfield
  {journal} {\bibinfo  {journal} {J. Chem. Phys.}\ }\textbf {\bibinfo {volume}
  {152}} (\bibinfo {year} {2020})}\BibitemShut {NoStop}%
\bibitem [{\citenamefont {Tilly}\ \emph {et~al.}(2022)\citenamefont {Tilly},
  \citenamefont {Chen}, \citenamefont {Cao}, \citenamefont {Picozzi},
  \citenamefont {Setia}, \citenamefont {Li}, \citenamefont {Grant},
  \citenamefont {Wossnig}, \citenamefont {Rungger}, \citenamefont {Booth} \emph
  {et~al.}}]{tilly2022variational}%
  \BibitemOpen
  \bibfield  {author} {\bibinfo {author} {\bibfnamefont {J.}~\bibnamefont
  {Tilly}}, \bibinfo {author} {\bibfnamefont {H.}~\bibnamefont {Chen}},
  \bibinfo {author} {\bibfnamefont {S.}~\bibnamefont {Cao}}, \bibinfo {author}
  {\bibfnamefont {D.}~\bibnamefont {Picozzi}}, \bibinfo {author} {\bibfnamefont
  {K.}~\bibnamefont {Setia}}, \bibinfo {author} {\bibfnamefont
  {Y.}~\bibnamefont {Li}}, \bibinfo {author} {\bibfnamefont {E.}~\bibnamefont
  {Grant}}, \bibinfo {author} {\bibfnamefont {L.}~\bibnamefont {Wossnig}},
  \bibinfo {author} {\bibfnamefont {I.}~\bibnamefont {Rungger}}, \bibinfo
  {author} {\bibfnamefont {G.~H.}\ \bibnamefont {Booth}},  \emph {et~al.},\
  }\href {https://www.sciencedirect.com/science/article/pii/S0370157322003118}
  {\bibfield  {journal} {\bibinfo  {journal} {Phys. Rep.}\ }\textbf {\bibinfo
  {volume} {986}},\ \bibinfo {pages} {1} (\bibinfo {year} {2022})}\BibitemShut
  {NoStop}%
\bibitem [{\citenamefont {Claudino}\ \emph {et~al.}(2020)\citenamefont
  {Claudino}, \citenamefont {Wright}, \citenamefont {McCaskey},\ and\
  \citenamefont {Humble}}]{claudino2020benchmarking}%
  \BibitemOpen
  \bibfield  {author} {\bibinfo {author} {\bibfnamefont {D.}~\bibnamefont
  {Claudino}}, \bibinfo {author} {\bibfnamefont {J.}~\bibnamefont {Wright}},
  \bibinfo {author} {\bibfnamefont {A.~J.}\ \bibnamefont {McCaskey}}, \ and\
  \bibinfo {author} {\bibfnamefont {T.~S.}\ \bibnamefont {Humble}},\ }\href
  {https://www.frontiersin.org/journals/chemistry/articles/10.3389/fchem.2020.606863/full}
  {\bibfield  {journal} {\bibinfo  {journal} {Front. Chem.}\ }\textbf {\bibinfo
  {volume} {8}},\ \bibinfo {pages} {606863} (\bibinfo {year}
  {2020})}\BibitemShut {NoStop}%
\bibitem [{\citenamefont {Gupta}\ \emph {et~al.}(2022)\citenamefont {Gupta},
  \citenamefont {Sajjan}, \citenamefont {Levine},\ and\ \citenamefont
  {Kais}}]{gupta2022variational}%
  \BibitemOpen
  \bibfield  {author} {\bibinfo {author} {\bibfnamefont {R.}~\bibnamefont
  {Gupta}}, \bibinfo {author} {\bibfnamefont {M.}~\bibnamefont {Sajjan}},
  \bibinfo {author} {\bibfnamefont {R.~D.}\ \bibnamefont {Levine}}, \ and\
  \bibinfo {author} {\bibfnamefont {S.}~\bibnamefont {Kais}},\ }\href
  {https://pubs.rsc.org/en/content/articlelanding/2022/cp/d2cp04493e/unauth}
  {\bibfield  {journal} {\bibinfo  {journal} {Phys. Chem. Chem. Phys.}\
  }\textbf {\bibinfo {volume} {24}},\ \bibinfo {pages} {28870} (\bibinfo {year}
  {2022})}\BibitemShut {NoStop}%
\bibitem [{\citenamefont {Miranowicz}\ \emph {et~al.}(2014)\citenamefont
  {Miranowicz}, \citenamefont {Bartkiewicz}, \citenamefont {Pe{\v{r}}ina~Jr},
  \citenamefont {Koashi}, \citenamefont {Imoto},\ and\ \citenamefont
  {Nori}}]{miranowicz2014optimal}%
  \BibitemOpen
  \bibfield  {author} {\bibinfo {author} {\bibfnamefont {A.}~\bibnamefont
  {Miranowicz}}, \bibinfo {author} {\bibfnamefont {K.}~\bibnamefont
  {Bartkiewicz}}, \bibinfo {author} {\bibfnamefont {J.}~\bibnamefont
  {Pe{\v{r}}ina~Jr}}, \bibinfo {author} {\bibfnamefont {M.}~\bibnamefont
  {Koashi}}, \bibinfo {author} {\bibfnamefont {N.}~\bibnamefont {Imoto}}, \
  and\ \bibinfo {author} {\bibfnamefont {F.}~\bibnamefont {Nori}},\ }\href
  {https://journals.aps.org/pra/abstract/10.1103/PhysRevA.90.062123} {\bibfield
   {journal} {\bibinfo  {journal} {Phys. Rev. A.}\ }\textbf {\bibinfo {volume}
  {90}},\ \bibinfo {pages} {062123} (\bibinfo {year} {2014})}\BibitemShut
  {NoStop}%
\bibitem [{\citenamefont {Melko}\ \emph {et~al.}(2019)\citenamefont {Melko},
  \citenamefont {Carleo}, \citenamefont {Carrasquilla},\ and\ \citenamefont
  {Cirac}}]{melko2019restricted}%
  \BibitemOpen
  \bibfield  {author} {\bibinfo {author} {\bibfnamefont {R.~G.}\ \bibnamefont
  {Melko}}, \bibinfo {author} {\bibfnamefont {G.}~\bibnamefont {Carleo}},
  \bibinfo {author} {\bibfnamefont {J.}~\bibnamefont {Carrasquilla}}, \ and\
  \bibinfo {author} {\bibfnamefont {J.~I.}\ \bibnamefont {Cirac}},\ }\href
  {https://www.nature.com/articles/s41567-019-0545-1} {\bibfield  {journal}
  {\bibinfo  {journal} {Nat. Phys.}\ }\textbf {\bibinfo {volume} {15}},\
  \bibinfo {pages} {887} (\bibinfo {year} {2019})}\BibitemShut {NoStop}%
\bibitem [{\citenamefont {Koutn{\`y}}\ \emph {et~al.}(2022)\citenamefont
  {Koutn{\`y}}, \citenamefont {Motka}, \citenamefont {Hradil}, \citenamefont
  {{\v{R}}eh{\'a}{\v{c}}ek},\ and\ \citenamefont
  {S{\'a}nchez-Soto}}]{koutny2022neural}%
  \BibitemOpen
  \bibfield  {author} {\bibinfo {author} {\bibfnamefont {D.}~\bibnamefont
  {Koutn{\`y}}}, \bibinfo {author} {\bibfnamefont {L.}~\bibnamefont {Motka}},
  \bibinfo {author} {\bibfnamefont {Z.}~\bibnamefont {Hradil}}, \bibinfo
  {author} {\bibfnamefont {J.}~\bibnamefont {{\v{R}}eh{\'a}{\v{c}}ek}}, \ and\
  \bibinfo {author} {\bibfnamefont {L.~L.}\ \bibnamefont {S{\'a}nchez-Soto}},\
  }\href {https://journals.aps.org/pra/abstract/10.1103/PhysRevA.106.012409}
  {\bibfield  {journal} {\bibinfo  {journal} {Phys. Rev. A.}\ }\textbf
  {\bibinfo {volume} {106}},\ \bibinfo {pages} {012409} (\bibinfo {year}
  {2022})}\BibitemShut {NoStop}%
\bibitem [{\citenamefont {Carrasquilla}\ \emph {et~al.}(2019)\citenamefont
  {Carrasquilla}, \citenamefont {Torlai}, \citenamefont {Melko},\ and\
  \citenamefont {Aolita}}]{carrasquilla2019reconstructing}%
  \BibitemOpen
  \bibfield  {author} {\bibinfo {author} {\bibfnamefont {J.}~\bibnamefont
  {Carrasquilla}}, \bibinfo {author} {\bibfnamefont {G.}~\bibnamefont
  {Torlai}}, \bibinfo {author} {\bibfnamefont {R.~G.}\ \bibnamefont {Melko}}, \
  and\ \bibinfo {author} {\bibfnamefont {L.}~\bibnamefont {Aolita}},\ }\href
  {https://www.nature.com/articles/s42256-019-0028-1} {\bibfield  {journal}
  {\bibinfo  {journal} {Nat. Mach. Intell.}\ }\textbf {\bibinfo {volume} {1}},\
  \bibinfo {pages} {155} (\bibinfo {year} {2019})}\BibitemShut {NoStop}%
\bibitem [{\citenamefont {Akhtar}\ \emph {et~al.}(2023)\citenamefont {Akhtar},
  \citenamefont {Hu},\ and\ \citenamefont {You}}]{akhtar2023scalable}%
  \BibitemOpen
  \bibfield  {author} {\bibinfo {author} {\bibfnamefont {A.~A.}\ \bibnamefont
  {Akhtar}}, \bibinfo {author} {\bibfnamefont {H.-Y.}\ \bibnamefont {Hu}}, \
  and\ \bibinfo {author} {\bibfnamefont {Y.-Z.}\ \bibnamefont {You}},\ }\href
  {https://quantum-journal.org/papers/q-2023-06-01-1026/} {\bibfield  {journal}
  {\bibinfo  {journal} {Quantum}\ }\textbf {\bibinfo {volume} {7}},\ \bibinfo
  {pages} {1026} (\bibinfo {year} {2023})}\BibitemShut {NoStop}%
\bibitem [{\citenamefont {Cotler}\ and\ \citenamefont
  {Wilczek}(2020)}]{cotler2020quantum}%
  \BibitemOpen
  \bibfield  {author} {\bibinfo {author} {\bibfnamefont {J.}~\bibnamefont
  {Cotler}}\ and\ \bibinfo {author} {\bibfnamefont {F.}~\bibnamefont
  {Wilczek}},\ }\href
  {https://journals.aps.org/prl/abstract/10.1103/PhysRevLett.124.100401}
  {\bibfield  {journal} {\bibinfo  {journal} {Phys. Rev. Lett.}\ }\textbf
  {\bibinfo {volume} {124}},\ \bibinfo {pages} {100401} (\bibinfo {year}
  {2020})}\BibitemShut {NoStop}%
\bibitem [{\citenamefont {Bonet-Monroig}\ \emph {et~al.}(2020)\citenamefont
  {Bonet-Monroig}, \citenamefont {Babbush},\ and\ \citenamefont
  {O’Brien}}]{bonet2020nearly}%
  \BibitemOpen
  \bibfield  {author} {\bibinfo {author} {\bibfnamefont {X.}~\bibnamefont
  {Bonet-Monroig}}, \bibinfo {author} {\bibfnamefont {R.}~\bibnamefont
  {Babbush}}, \ and\ \bibinfo {author} {\bibfnamefont {T.~E.}\ \bibnamefont
  {O’Brien}},\ }\href
  {https://journals.aps.org/prx/abstract/10.1103/PhysRevX.10.031064} {\bibfield
   {journal} {\bibinfo  {journal} {Phys. Rev. X.}\ }\textbf {\bibinfo {volume}
  {10}},\ \bibinfo {pages} {031064} (\bibinfo {year} {2020})}\BibitemShut
  {NoStop}%
\bibitem [{\citenamefont {Garc{\'\i}a-P{\'e}rez}\ \emph
  {et~al.}(2020)\citenamefont {Garc{\'\i}a-P{\'e}rez}, \citenamefont {Rossi},
  \citenamefont {Sokolov}, \citenamefont {Borrelli},\ and\ \citenamefont
  {Maniscalco}}]{garcia2020pairwise}%
  \BibitemOpen
  \bibfield  {author} {\bibinfo {author} {\bibfnamefont {G.}~\bibnamefont
  {Garc{\'\i}a-P{\'e}rez}}, \bibinfo {author} {\bibfnamefont {M.~A.}\
  \bibnamefont {Rossi}}, \bibinfo {author} {\bibfnamefont {B.}~\bibnamefont
  {Sokolov}}, \bibinfo {author} {\bibfnamefont {E.-M.}\ \bibnamefont
  {Borrelli}}, \ and\ \bibinfo {author} {\bibfnamefont {S.}~\bibnamefont
  {Maniscalco}},\ }\href
  {https://journals.aps.org/prresearch/abstract/10.1103/PhysRevResearch.2.023393}
  {\bibfield  {journal} {\bibinfo  {journal} {Phys. Rev. Res.}\ }\textbf
  {\bibinfo {volume} {2}},\ \bibinfo {pages} {023393} (\bibinfo {year}
  {2020})}\BibitemShut {NoStop}%
\bibitem [{\citenamefont {Yang}\ \emph {et~al.}(2023)\citenamefont {Yang},
  \citenamefont {Ru}, \citenamefont {Cao}, \citenamefont {Zheludev},\ and\
  \citenamefont {Gao}}]{yang2023experimental}%
  \BibitemOpen
  \bibfield  {author} {\bibinfo {author} {\bibfnamefont {Z.}~\bibnamefont
  {Yang}}, \bibinfo {author} {\bibfnamefont {S.}~\bibnamefont {Ru}}, \bibinfo
  {author} {\bibfnamefont {L.}~\bibnamefont {Cao}}, \bibinfo {author}
  {\bibfnamefont {N.}~\bibnamefont {Zheludev}}, \ and\ \bibinfo {author}
  {\bibfnamefont {W.}~\bibnamefont {Gao}},\ }\href
  {https://journals.aps.org/prl/abstract/10.1103/PhysRevLett.130.050804}
  {\bibfield  {journal} {\bibinfo  {journal} {Phys. Rev. Lett.}\ }\textbf
  {\bibinfo {volume} {130}},\ \bibinfo {pages} {050804} (\bibinfo {year}
  {2023})}\BibitemShut {NoStop}%
\bibitem [{\citenamefont {Hu}\ \emph {et~al.}(2024)\citenamefont {Hu},
  \citenamefont {Wei}, \citenamefont {Liu}, \citenamefont {Che}, \citenamefont
  {Zhou}, \citenamefont {Xie}, \citenamefont {Qin}, \citenamefont {Hu},
  \citenamefont {Yuan}, \citenamefont {Zhou} \emph
  {et~al.}}]{hu2024experimental}%
  \BibitemOpen
  \bibfield  {author} {\bibinfo {author} {\bibfnamefont {C.-K.}\ \bibnamefont
  {Hu}}, \bibinfo {author} {\bibfnamefont {C.}~\bibnamefont {Wei}}, \bibinfo
  {author} {\bibfnamefont {C.}~\bibnamefont {Liu}}, \bibinfo {author}
  {\bibfnamefont {L.}~\bibnamefont {Che}}, \bibinfo {author} {\bibfnamefont
  {Y.}~\bibnamefont {Zhou}}, \bibinfo {author} {\bibfnamefont {G.}~\bibnamefont
  {Xie}}, \bibinfo {author} {\bibfnamefont {H.}~\bibnamefont {Qin}}, \bibinfo
  {author} {\bibfnamefont {G.}~\bibnamefont {Hu}}, \bibinfo {author}
  {\bibfnamefont {H.}~\bibnamefont {Yuan}}, \bibinfo {author} {\bibfnamefont
  {R.}~\bibnamefont {Zhou}},  \emph {et~al.},\ }\href
  {https://arxiv.org/abs/2409.12614} {\bibfield  {journal} {\bibinfo  {journal}
  {arXiv:2409.12614}\ } (\bibinfo {year} {2024})}\BibitemShut {NoStop}%
\bibitem [{\citenamefont {Gokhale}\ \emph {et~al.}(2019)\citenamefont
  {Gokhale}, \citenamefont {Angiuli}, \citenamefont {Ding}, \citenamefont
  {Gui}, \citenamefont {Tomesh}, \citenamefont {Suchara}, \citenamefont
  {Martonosi},\ and\ \citenamefont {Chong}}]{gokhale2019minimizing}%
  \BibitemOpen
  \bibfield  {author} {\bibinfo {author} {\bibfnamefont {P.}~\bibnamefont
  {Gokhale}}, \bibinfo {author} {\bibfnamefont {O.}~\bibnamefont {Angiuli}},
  \bibinfo {author} {\bibfnamefont {Y.}~\bibnamefont {Ding}}, \bibinfo {author}
  {\bibfnamefont {K.}~\bibnamefont {Gui}}, \bibinfo {author} {\bibfnamefont
  {T.}~\bibnamefont {Tomesh}}, \bibinfo {author} {\bibfnamefont
  {M.}~\bibnamefont {Suchara}}, \bibinfo {author} {\bibfnamefont
  {M.}~\bibnamefont {Martonosi}}, \ and\ \bibinfo {author} {\bibfnamefont
  {F.~T.}\ \bibnamefont {Chong}},\ }\href {https://arxiv.org/abs/1907.13623}
  {\bibfield  {journal} {\bibinfo  {journal} {arXiv:1907.13623}\ } (\bibinfo
  {year} {2019})}\BibitemShut {NoStop}%
\bibitem [{\citenamefont {Yen}\ \emph {et~al.}(2020)\citenamefont {Yen},
  \citenamefont {Verteletskyi},\ and\ \citenamefont
  {Izmaylov}}]{yen2020measuring}%
  \BibitemOpen
  \bibfield  {author} {\bibinfo {author} {\bibfnamefont {T.-C.}\ \bibnamefont
  {Yen}}, \bibinfo {author} {\bibfnamefont {V.}~\bibnamefont {Verteletskyi}}, \
  and\ \bibinfo {author} {\bibfnamefont {A.~F.}\ \bibnamefont {Izmaylov}},\
  }\href {https://pubs.acs.org/doi/abs/10.1021/acs.jctc.0c00008} {\bibfield
  {journal} {\bibinfo  {journal} {J. Chem. Theory Comput.}\ }\textbf {\bibinfo
  {volume} {16}},\ \bibinfo {pages} {2400} (\bibinfo {year}
  {2020})}\BibitemShut {NoStop}%
\bibitem [{\citenamefont {Crawford}\ \emph {et~al.}(2021)\citenamefont
  {Crawford}, \citenamefont {van Straaten}, \citenamefont {Wang}, \citenamefont
  {Parks}, \citenamefont {Campbell},\ and\ \citenamefont
  {Brierley}}]{crawford2021efficient}%
  \BibitemOpen
  \bibfield  {author} {\bibinfo {author} {\bibfnamefont {O.}~\bibnamefont
  {Crawford}}, \bibinfo {author} {\bibfnamefont {B.}~\bibnamefont {van
  Straaten}}, \bibinfo {author} {\bibfnamefont {D.}~\bibnamefont {Wang}},
  \bibinfo {author} {\bibfnamefont {T.}~\bibnamefont {Parks}}, \bibinfo
  {author} {\bibfnamefont {E.}~\bibnamefont {Campbell}}, \ and\ \bibinfo
  {author} {\bibfnamefont {S.}~\bibnamefont {Brierley}},\ }\href
  {https://quantum-journal.org/papers/q-2021-01-20-385/} {\bibfield  {journal}
  {\bibinfo  {journal} {Quantum}\ }\textbf {\bibinfo {volume} {5}},\ \bibinfo
  {pages} {385} (\bibinfo {year} {2021})}\BibitemShut {NoStop}%
\bibitem [{\citenamefont {Guo}\ and\ \citenamefont
  {Yang}(2023)}]{guo2023scalable}%
  \BibitemOpen
  \bibfield  {author} {\bibinfo {author} {\bibfnamefont {Y.}~\bibnamefont
  {Guo}}\ and\ \bibinfo {author} {\bibfnamefont {S.}~\bibnamefont {Yang}},\
  }\href {https://arxiv.org/abs/2307.16381} {\bibfield  {journal} {\bibinfo
  {journal} {arXiv:2307.16381}\ } (\bibinfo {year} {2023})}\BibitemShut
  {NoStop}%
\bibitem [{\citenamefont {Cao}\ \emph {et~al.}(2023)\citenamefont {Cao},
  \citenamefont {Wu}, \citenamefont {Chen}, \citenamefont {Gong}, \citenamefont
  {Wu}, \citenamefont {Ye}, \citenamefont {Zha}, \citenamefont {Qian},
  \citenamefont {Ying}, \citenamefont {Guo} \emph
  {et~al.}}]{cao2023generation}%
  \BibitemOpen
  \bibfield  {author} {\bibinfo {author} {\bibfnamefont {S.}~\bibnamefont
  {Cao}}, \bibinfo {author} {\bibfnamefont {B.}~\bibnamefont {Wu}}, \bibinfo
  {author} {\bibfnamefont {F.}~\bibnamefont {Chen}}, \bibinfo {author}
  {\bibfnamefont {M.}~\bibnamefont {Gong}}, \bibinfo {author} {\bibfnamefont
  {Y.}~\bibnamefont {Wu}}, \bibinfo {author} {\bibfnamefont {Y.}~\bibnamefont
  {Ye}}, \bibinfo {author} {\bibfnamefont {C.}~\bibnamefont {Zha}}, \bibinfo
  {author} {\bibfnamefont {H.}~\bibnamefont {Qian}}, \bibinfo {author}
  {\bibfnamefont {C.}~\bibnamefont {Ying}}, \bibinfo {author} {\bibfnamefont
  {S.}~\bibnamefont {Guo}},  \emph {et~al.},\ }\href
  {https://www.nature.com/articles/s41586-023-06195-1} {\bibfield  {journal}
  {\bibinfo  {journal} {Nature}\ }\textbf {\bibinfo {volume} {619}},\ \bibinfo
  {pages} {738} (\bibinfo {year} {2023})}\BibitemShut {NoStop}%
\bibitem [{sm()}]{sm}%
  \BibitemOpen
  \href@noop {} {\bibinfo  {journal} {See the supplemental information for more
  details}\ }\BibitemShut {NoStop}%
\bibitem [{\citenamefont {Lu}\ \emph {et~al.}(2016)\citenamefont {Lu},
  \citenamefont {Xin}, \citenamefont {Yu}, \citenamefont {Ji}, \citenamefont
  {Chen}, \citenamefont {Long}, \citenamefont {Baugh}, \citenamefont {Peng},
  \citenamefont {Zeng},\ and\ \citenamefont {Laflamme}}]{lu2016tomography}%
  \BibitemOpen
\bibfield  {journal} {  }\bibfield  {author} {\bibinfo {author} {\bibfnamefont
  {D.}~\bibnamefont {Lu}}, \bibinfo {author} {\bibfnamefont {T.}~\bibnamefont
  {Xin}}, \bibinfo {author} {\bibfnamefont {N.}~\bibnamefont {Yu}}, \bibinfo
  {author} {\bibfnamefont {Z.}~\bibnamefont {Ji}}, \bibinfo {author}
  {\bibfnamefont {J.}~\bibnamefont {Chen}}, \bibinfo {author} {\bibfnamefont
  {G.}~\bibnamefont {Long}}, \bibinfo {author} {\bibfnamefont {J.}~\bibnamefont
  {Baugh}}, \bibinfo {author} {\bibfnamefont {X.}~\bibnamefont {Peng}},
  \bibinfo {author} {\bibfnamefont {B.}~\bibnamefont {Zeng}}, \ and\ \bibinfo
  {author} {\bibfnamefont {R.}~\bibnamefont {Laflamme}},\ }\href
  {https://journals.aps.org/prl/abstract/10.1103/PhysRevLett.116.230501}
  {\bibfield  {journal} {\bibinfo  {journal} {Phys. Rev. Lett.}\ }\textbf
  {\bibinfo {volume} {116}},\ \bibinfo {pages} {230501} (\bibinfo {year}
  {2016})}\BibitemShut {NoStop}%
\bibitem [{\citenamefont {Verstraete}\ \emph {et~al.}(2002)\citenamefont
  {Verstraete}, \citenamefont {Dehaene}, \citenamefont {De~Moor},\ and\
  \citenamefont {Verschelde}}]{verstraete2002four}%
  \BibitemOpen
  \bibfield  {author} {\bibinfo {author} {\bibfnamefont {F.}~\bibnamefont
  {Verstraete}}, \bibinfo {author} {\bibfnamefont {J.}~\bibnamefont {Dehaene}},
  \bibinfo {author} {\bibfnamefont {B.}~\bibnamefont {De~Moor}}, \ and\
  \bibinfo {author} {\bibfnamefont {H.}~\bibnamefont {Verschelde}},\ }\href
  {https://journals.aps.org/pra/abstract/10.1103/PhysRevA.65.052112} {\bibfield
   {journal} {\bibinfo  {journal} {Phys. Rev. A.}\ }\textbf {\bibinfo {volume}
  {65}},\ \bibinfo {pages} {052112} (\bibinfo {year} {2002})}\BibitemShut
  {NoStop}%
\bibitem [{\citenamefont {Giordano}\ and\ \citenamefont
  {Martin-Delgado}(2022)}]{giordano2022reinforcement}%
  \BibitemOpen
  \bibfield  {author} {\bibinfo {author} {\bibfnamefont {S.}~\bibnamefont
  {Giordano}}\ and\ \bibinfo {author} {\bibfnamefont {M.~A.}\ \bibnamefont
  {Martin-Delgado}},\ }\href
  {https://journals.aps.org/prresearch/abstract/10.1103/PhysRevResearch.4.043056}
  {\bibfield  {journal} {\bibinfo  {journal} {Phys. Rev. Res.}\ }\textbf
  {\bibinfo {volume} {4}},\ \bibinfo {pages} {043056} (\bibinfo {year}
  {2022})}\BibitemShut {NoStop}%
\bibitem [{\citenamefont {Vintskevich}\ \emph {et~al.}(2023)\citenamefont
  {Vintskevich}, \citenamefont {Bao}, \citenamefont {Nomerotski}, \citenamefont
  {Stankus},\ and\ \citenamefont {Grigoriev}}]{vintskevich2023classification}%
  \BibitemOpen
  \bibfield  {author} {\bibinfo {author} {\bibfnamefont {S.}~\bibnamefont
  {Vintskevich}}, \bibinfo {author} {\bibfnamefont {N.}~\bibnamefont {Bao}},
  \bibinfo {author} {\bibfnamefont {A.}~\bibnamefont {Nomerotski}}, \bibinfo
  {author} {\bibfnamefont {P.}~\bibnamefont {Stankus}}, \ and\ \bibinfo
  {author} {\bibfnamefont {D.}~\bibnamefont {Grigoriev}},\ }\href
  {https://journals.aps.org/pra/abstract/10.1103/PhysRevA.107.032421}
  {\bibfield  {journal} {\bibinfo  {journal} {Phys. Rev. A.}\ }\textbf
  {\bibinfo {volume} {107}},\ \bibinfo {pages} {032421} (\bibinfo {year}
  {2023})}\BibitemShut {NoStop}%
\bibitem [{\citenamefont {Khaneja}\ \emph {et~al.}(2005)\citenamefont
  {Khaneja}, \citenamefont {Reiss}, \citenamefont {Kehlet}, \citenamefont
  {Schulte-Herbr{\"u}ggen},\ and\ \citenamefont {Glaser}}]{khaneja2005optimal}%
  \BibitemOpen
  \bibfield  {author} {\bibinfo {author} {\bibfnamefont {N.}~\bibnamefont
  {Khaneja}}, \bibinfo {author} {\bibfnamefont {T.}~\bibnamefont {Reiss}},
  \bibinfo {author} {\bibfnamefont {C.}~\bibnamefont {Kehlet}}, \bibinfo
  {author} {\bibfnamefont {T.}~\bibnamefont {Schulte-Herbr{\"u}ggen}}, \ and\
  \bibinfo {author} {\bibfnamefont {S.~J.}\ \bibnamefont {Glaser}},\ }\href
  {https://www.sciencedirect.com/science/article/pii/S1090780704003696}
  {\bibfield  {journal} {\bibinfo  {journal} {J. Magn. Reson.}\ }\textbf
  {\bibinfo {volume} {172}},\ \bibinfo {pages} {296} (\bibinfo {year}
  {2005})}\BibitemShut {NoStop}%
\bibitem [{\citenamefont {Wootters}(2001)}]{wootters2001entanglement}%
  \BibitemOpen
  \bibfield  {author} {\bibinfo {author} {\bibfnamefont {W.~K.}\ \bibnamefont
  {Wootters}},\ }\href {https://www.rintonpress.com/journals/qic-1-1/eof2.pdf}
  {\bibfield  {journal} {\bibinfo  {journal} {Quantum Inf. Comput.}\ }\textbf
  {\bibinfo {volume} {1}},\ \bibinfo {pages} {27} (\bibinfo {year}
  {2001})}\BibitemShut {NoStop}%
\bibitem [{\citenamefont {Phillips}(2012)}]{phillips2012chernoff}%
  \BibitemOpen
  \bibfield  {author} {\bibinfo {author} {\bibfnamefont {J.~M.}\ \bibnamefont
  {Phillips}},\ }\href {https://arxiv.org/abs/1209.6396} {\bibfield  {journal}
  {\bibinfo  {journal} {arXiv:1209.6396}\ } (\bibinfo {year}
  {2012})}\BibitemShut {NoStop}%
\bibitem [{\citenamefont {Chang}\ \emph {et~al.}(2020)\citenamefont {Chang},
  \citenamefont {Jones}, \citenamefont {Yao}, \citenamefont {Graf},\ and\
  \citenamefont {Jain}}]{chang2020hybrid}%
  \BibitemOpen
  \bibfield  {author} {\bibinfo {author} {\bibfnamefont {C.-Y.}\ \bibnamefont
  {Chang}}, \bibinfo {author} {\bibfnamefont {E.}~\bibnamefont {Jones}},
  \bibinfo {author} {\bibfnamefont {Y.}~\bibnamefont {Yao}}, \bibinfo {author}
  {\bibfnamefont {P.}~\bibnamefont {Graf}}, \ and\ \bibinfo {author}
  {\bibfnamefont {R.}~\bibnamefont {Jain}},\ }\href
  {https://arxiv.org/abs/2010.07852} {\bibfield  {journal} {\bibinfo  {journal}
  {arXiv:2010.07852}\ } (\bibinfo {year} {2020})}\BibitemShut {NoStop}%
\bibitem [{\citenamefont {Hansenne}\ \emph {et~al.}(2024)\citenamefont
  {Hansenne}, \citenamefont {Qu}, \citenamefont {Weinbrenner}, \citenamefont
  {de~Gois}, \citenamefont {Wang}, \citenamefont {Ming}, \citenamefont {Yang},
  \citenamefont {Horodecki}, \citenamefont {Gao},\ and\ \citenamefont
  {G{\"u}hne}}]{hansenne2024optimal}%
  \BibitemOpen
  \bibfield  {author} {\bibinfo {author} {\bibfnamefont {K.}~\bibnamefont
  {Hansenne}}, \bibinfo {author} {\bibfnamefont {R.}~\bibnamefont {Qu}},
  \bibinfo {author} {\bibfnamefont {L.~T.}\ \bibnamefont {Weinbrenner}},
  \bibinfo {author} {\bibfnamefont {C.}~\bibnamefont {de~Gois}}, \bibinfo
  {author} {\bibfnamefont {H.}~\bibnamefont {Wang}}, \bibinfo {author}
  {\bibfnamefont {Y.}~\bibnamefont {Ming}}, \bibinfo {author} {\bibfnamefont
  {Z.}~\bibnamefont {Yang}}, \bibinfo {author} {\bibfnamefont {P.}~\bibnamefont
  {Horodecki}}, \bibinfo {author} {\bibfnamefont {W.}~\bibnamefont {Gao}}, \
  and\ \bibinfo {author} {\bibfnamefont {O.}~\bibnamefont {G{\"u}hne}},\ }\href
  {https://arxiv.org/abs/2408.05730} {\bibfield  {journal} {\bibinfo  {journal}
  {arXiv:2408.05730}\ } (\bibinfo {year} {2024})}\BibitemShut {NoStop}%
\end{thebibliography}

%

\end{document}